\documentclass[lettersize,journal]{IEEEtran}
\usepackage{amsmath, amsfonts, mathrsfs, amsthm}
\usepackage{algorithm, algpseudocode}
\usepackage{bbm}
\usepackage{array}
\usepackage{acronym} 
\usepackage[caption=false,font=normalsize,labelfont=sf,textfont=sf]{subfig}
\usepackage{textcomp}
\usepackage{stfloats}
\usepackage{setspace} 
\usepackage{url}
\usepackage{verbatim}
\usepackage{graphicx}
\usepackage{booktabs}   
\usepackage{multirow}
\usepackage{makecell}
\usepackage{siunitx}
\usepackage{subcaption}
\usepackage{subfig}
\usepackage{orcidlink}
\usepackage{tikz}
\usepackage{adjustbox}
\usepackage[table]{xcolor}
\usepackage{colortbl}
\usetikzlibrary{arrows.meta,positioning, calc}
\usepackage[normalem]{ulem}
\usepackage{array}
\usepackage{enumitem}
\newcolumntype{M}[1]{>{\centering\arraybackslash}m{#1}} 
\newlength{\panelheight}
\usepackage[backend=biber,style=ieee]{biblatex} 
\usepackage{pifont}
\newcommand{\cmark}{\ding{51}} 
\newcommand{\xmark}{\ding{55}} 

\newtheorem{theorem}{Theorem}

\acrodef{RIR}{Room Impulse Response}
\acrodef{FIR}{Finite Impulse Response}
\acrodef{PSD}{Power Spectral Density}
\acrodef{CPSD}{Cross Power Spectral Density}
\acrodef{UQ}{Uncertainty Quantification}
\acrodef{SSL}{Sound Source Localization}
\acrodef{CP}{Conformal Prediction}
\acrodef{STFT}{Short-Time Fourier Transform}
\acrodef{GCC-PHAT}{Generalized Cross-Correlation with Phase Transform}
\acrodef{SRP-PHAT}{Steered-Response Power with Phase Transform}
\acrodef{SRP}{Steered-Response Power}
\acrodef{IPD}{Inter-Channel Phase Difference}
\acrodef{DNN}{Deep Neural Network}
\acrodef{TDOA}{Time-Difference of Arrival}
\acrodef{CRC}{Conformal Risk Control}
\acrodef{RCP}{Risk-Controlling Prediction}
\acrodef{RCPS}{Risk-Controlling Prediction Set}
\acrodef{FNR}{False Negative Rate}
\acrodef{LTT}{Learn Then Test}
\acrodef{DOA}{Direction of Arrival}
\acrodef{FWER}{Family-Wise Error Rate}
\acrodef{FST}{Fixed Sequence Testing}
\acrodef{SGT}{Sequential Graphical Testing}
\acrodef{WSR}{Waudby-Smith–Ramdas}
\acrodef{HB}{Hoeffding–Bentkus}
\acrodef{MOO}{Multi-Objective Optimization}
\acrodef{ROI}{Region of Interest}
\acrodef{AWGN}{Additive White Gaussian Noise}
\acrodef{SNR}{Signal-to-Noise Ratio}
\acrodef{MC}{Mis-Coverage}
\acrodef{NN}{neural network}
\acrodef{MD}{Mis-Detection}
\acrodef{FA}{False Alarm}
\acrodef{CRNN}{Convolutional Recurrent Neural Network}
\acrodef{MSE}{mean squared error}
\acrodef{SRP-DNN}{SRP-Deep Neural Network}
\acrodef{ESPRIT}{Estimation of Signal Parameters via Rotational Invariance Techniques}
\acrodef{MUSIC}{Multiple Signal Classification}
\acrodef{PI}{Prediction Interval}
\acrodef{T-F}{time-frequency}
\acrodef{MHT}{Multiple Hypothesis Testing}
\acrodef{PA}{Prediction Area}
\acrodef{EM}{Expectation Maximization}
\acrodef{LCMV}{Linear Constrained Minimum Variance}
\acrodef{MDN}{Mixture Density Network}
\acrodef{MAE}{Mean Angular Error}
\acrodef{SI-SDR}{Scale-Invariant Signal-to-Distortion Ratio}
\acrodef{ESTOI}{Extended Short-Time Objective Intelligibility}
\acrodef{PD}{Peak Detection}
\acrodef{CNN}{Convolutional Neural Network}

\newcommand{\BL}[1]{\textcolor{black}{#1}}

\begin{document}

\title{Uncertainty Quantification and Risk Control for\\ Multiple Sound Source Localization}

\author{Vadim Rozenfeld\,\orcidlink{0009-0004-0327-4581}, Bracha Laufer Goldshtein\, \orcidlink{0000-0003-2606-4793}
    
\thanks{The authors are with The School of Electrical \&  Computer Engineering, Tel-Aviv University, Tel-Aviv 6139001, Israel
(e-mail: vadimroz@mail.tau.ac.il;  blaufer@tauex.tau.ac.il). \BL{This work was supported in part by the Israel Science Foundation (Grant 2094/25), the Israeli
Ministry of Innovation, Science and Technology (Grant 1001818518)} and the Israel Innovation Authority’s
Magneton program (Grant 6602).}
}

\markboth{Journal of \LaTeX\ Class Files,~Vol.~14, No.~8, August~2021}
{Shell \MakeLowercase{\mathrmit{et al.}}: A Sample Article Using IEEEtran.cls for IEEE Journals}

\maketitle

\begin{abstract}
Reliable~\ac{SSL} plays an essential role in many downstream tasks, where informed decision-making depends not only on accurate localization but also on the confidence in each estimate. This need for reliability becomes even more pronounced in challenging conditions, such as reverberant environments and multi-source scenarios. However, existing~\ac{SSL} methods typically provide only point estimates, offering limited or no~\ac{UQ}.
We leverage the~\ac{CP} framework and its extensions for controlling general risk functions to develop two complementary~\ac{UQ} approaches for~\ac{SSL}. The first assumes that the number of active sources is known and constructs prediction regions that cover the true source locations. The second addresses the more challenging setting where the source count is unknown, by first reliably estimating the number of active sources and then forming corresponding prediction regions.
We evaluate the proposed methods on extensive simulations and real-world recordings across varying reverberation levels and source configurations. Results demonstrate reliable finite-sample guarantees and consistent performance for both known- and unknown-source-count scenarios, highlighting the practical utility of the proposed frameworks for uncertainty-aware SSL.
\end{abstract}


\section{Introduction}
\IEEEPARstart{A}{ccurate}~\acf{SSL} is a fundamental component in a wide range of acoustic and audio signal processing systems, including robot audition~\cite{rascon2017localization}, camera steering~\cite{benesty2000adaptive}, and speech enhancement and separation~\cite{mandel2010model}. In these applications, determining the~\ac{DOA} or positions of one or more active sources enables downstream modules such as beamforming, source separation, diarization, or navigation, to operate reliably. The demand for robust~\ac{SSL} has only intensified as systems encounter increasingly complex acoustic environments with reverberation, noise, and multiple sources.\looseness=-1

A large body of work has focused on developing algorithms capable of estimating~\ac{SSL} under challenging acoustic conditions. 
Several classical approaches generate likelihood maps over a predefined spatial grid. These include~\ac{SRP-PHAT}~\cite{dibiase2001robust} and~\ac{EM}-based methods~\cite{schwartz2013speaker}, which exploit~\acp{TDOA} between microphone pairs, as well as subspace-based techniques such as~\ac{MUSIC}~\cite{schmidt1986multiple} and~\ac{ESPRIT}~\cite{roy1989esprit}.
More recently,~\ac{DNN}-based models have been proposed and often demonstrate improved performance by learning complex patterns that are difficult to capture with traditional signal processing techniques~\cite{grumiaux2022survey}. Many of these approaches produce spatial likelihood maps by imitating classical localization methods~\cite{9287466,srp_dnn, neural-SRP}, or by formulating the~\ac{SSL} problem as a classification task over a set of classes obtained by discretizing the region of interest~\cite{shaybet2025srp, Chakrabarty2019, uncertainty_est_for_sll_dl}.\looseness=-1

Despite notable progress, most~\ac{SSL} systems \textcolor{black}{operate in a strict point-estimation manner}, meaning they output a single estimated location for each detected source, without quantifying the reliability of these estimates. However, errors are unavoidable since noise, reverberation, and interfering signals fundamentally distort the spatial cues on which SSL methods rely. Without a principled measure of uncertainty, downstream components must either trust the~\ac{SSL} output blindly or rely on heuristic thresholds that provide no statistical guarantees.\looseness=-1

A few recent approaches have proposed extracting reliability measures for~\ac{SSL} estimates to detect low-confidence estimates and out-of-distribution samples~\cite{shaybet2025srp,uncertainty_est_for_sll_dl}. In~\cite{shaybet2025srp}, Gaussian label smoothing is used during training, and the predicted peak value is treated as a proxy for reliability, while~\cite{uncertainty_est_for_sll_dl} models output uncertainty via a Dirichlet distribution. Both approaches provide useful reliability indicators, albeit without formal statistical guarantees.\looseness=-1

A widely used~\ac{UQ} approach that provides distribution-free finite-sample guarantees is~\ac{CP}~\cite{gammerman1998transduction, vovk2005algorithmic, angelopoulos2024theoretical}, which constructs~\acp{PI} that contain the true value with a user-specified probability.
A few recent attempts were made to apply~\acf{CP} for~\ac{SSL}. For the single-source case,~\cite{khurjekar2023uncertainty} applied~\ac{CP} using several uncertainty surrogates, including Monte-Carlo dropout, model ensembles, and quantile regression. Our prior work combined manifold-based~\ac{SSL} via Gaussian process regression~\cite{laufer2016multiple} with~\ac{CP} to efficiently construct statistically valid~\acp{PI}~\cite{CP_MMGP}. This kernel-based localization approach was replaced by a graph neural network in~\cite{gnn_ssl}, while constructing valid~\acp{PI} using an alternative Jackknife+ formulation~\cite{barber2021predictive}. In the multi-source setting,~\cite{khurjekar2024multi} assumes a known number of sources and models the conditional multi-source~\ac{DOA} distribution using a Gaussian mixture model parameterized by a~\ac{NN}, and~\ac{CP} is then applied to obtain~\acp{PI} from its outputs.\looseness=-1

These previous works~\cite{khurjekar2023uncertainty, CP_MMGP, gnn_ssl, khurjekar2024multi} heavily rely on classical~\ac{CP}, which is tailored exclusively to the binarized~\ac{MC} loss, namely, the probability that the true value lies outside the~\ac{PI}.
Yet, multi-source setups introduce additional and potentially competing losses beyond~\ac{MC}. For instance, errors may arise from failing to detect an existing speaker at all, or from declaring a speaker that does not exist. 
Moreover, classical~\ac{CP} is suited to calibrating only a single hyperparameter, such as a detection threshold, whereas practical systems involve multiple hyperparameters that must be tuned jointly. In addition, prior~\ac{UQ} frameworks are typically optimized for specific localization methods and do not generalize readily to alternative approaches.\looseness=-1

This work addresses these gaps by leveraging recent advances in the~\ac{CP} literature, enabling principled~\ac{UQ} for complex, multi-objective settings that classical~\ac{CP} cannot accommodate. We propose a general~\ac{UQ} framework applicable to any multi-source localization method that produces a likelihood map over a grid of candidate positions. We consider both scenarios in which the number of sources is known or unknown. 
The key idea is to associate each position estimate with an interpretable confidence information, expressed as prediction region, on a 2D spatial likelihood map that contains the true source location with high probability. When the number of sources is unknown, the framework further enables controlled detection of the correct number of sources. Extensive experimental studies demonstrate the effectiveness of the proposed framework across a wide range of conditions.\looseness=-1

Our main contributions are summarized as follows: 
\begin{itemize}[leftmargin=*]
    \item Our proposed frameworks operate on spatial likelihood maps from \textit{any} classical or~\ac{DNN}-based localization method, yielding finite-sample, statistically valid prediction regions containing the true source location with high probability.
    \item We introduce a method for constructing contiguous prediction regions around each \ac{SSL} estimate that adapt to the local likelihood landscape, without imposing any geometric or distributional assumptions on the region shape. The size of this region directly quantifies uncertainty, with larger regions indicating lower confidence and vice versa. Such information can naturally inform downstream decision-making.
    \item For the unknown-source case, beyond classical~\ac{MC} risk  control, we additionally provide principled simultaneous control over the risk of failing to detect an existing speaker, while discouraging spurious detections and minimizing the size of the prediction region.
    \item We evaluate the proposed frameworks on single- and multi-source tasks with both known and unknown numbers of sources. We consider both classical and \ac{DNN}-based likelihood maps obtained for simulated environments, and real-world recordings, covering both static- and moving-source scenarios. Our code is publicly available on GitHub.\looseness=-1\footnote{ \url{https://github.com/vadimroz/UQ_in_multi_SSL}}
\end{itemize}

The remainder of the paper is organized as follows.
Section~\ref{sec:problem_formulation} formulates the multi-source localization problem.
Section~\ref{sec:background} reviews~\ac{SSL} methods based on constructing spatial likelihood maps.
Section~\ref{sec:conformal_prediction} provides background on \ac{CP} and risk control extensions.
Section~\ref{sec:proposed_method} presents the proposed approaches for risk control in settings with known and unknown numbers of sources.
Section~\ref{sec:simulation_and_results} describes experimental setup and results.
Section~\ref{sec:limiations_future_work} describes limitations and future work, and Section~\ref{sec:conclusion} concludes the paper.\looseness=-1

\section{Problem Formulation}
\label{sec:problem_formulation}
We consider a reverberant acoustic enclosure with $K$ static sources, $1 \le K \le K_{\max}$, where $K$ is unknown but $K_{\max}$ is known.
Let~$\{\mathbf{p}^*_k\}_{k=1}^{K}$ denote the 2D~\acp{DOA} of the sources, where $\mathbf{p}_k^*\! =\![\phi_k,\theta_k]^{\mathsf{T}}\in\mathbb{R}^2$, 
with~$\phi_k\! \in \![0,\pi]$ and~$\theta_k\!\in\![-\pi,\pi]$ denoting the elevation and azimuth angles, respectively.
The signal received at the~$m$-th microphone in the~\ac{STFT} domain is expressed as\looseness=-1
\begin{equation}
    X_m(t,f)
    = \sum_{k=1}^{K} A_m(f,\mathbf{p}_k^*)\, S_k(t,f)
    + U_m(t,f),
    \label{eq:problem_formulation_in_freq}
\end{equation}
where~$t\!\in\!\{1,\ldots, T\}$ and~$f\!\in\!\{1, \ldots, F\}$ denote the frame and frequency indices, respectively.
Here, $S_k(t,f)$ is the $k$-th source signal, $A_m(f,\mathbf{p}_k^*)$ is the~\ac{RIR} relating the $k$-th source at $\mathbf{p}^*_k$ and the $m$-th microphone, and $U_m(t,f)$ is an additive noise. \looseness=-1

Our framework builds on existing \ac{SSL} methods that produce 2D spatial likelihood maps, from which the \acp{DOA} are estimated. We consider both classical approaches, such as \ac{SRP-PHAT}~\cite{dibiase2001robust}, and neural network–based models~\cite{srp_dnn,neural-SRP,9287466,shaybet2025srp,Chakrabarty2019}. Given a generic spatial likelihood map, our goal is to enable principled \ac{UQ} and risk control by constructing prediction regions that cover the true source locations with high probability. When the number of sources is unknown, we additionally wish to control the error of incorrectly estimating the source count.\looseness=-1

\section{Background on Sound Source Localization}
\label{sec:background}
This section reviews two representative localization methods that produce 2D spatial likelihood maps: the classical \ac{SRP-PHAT} method~\cite{dibiase2001robust} and the learning-based SRP-DNN~\cite{srp_dnn}. \textcolor{black}{SRP-like methods are suitable in this context as they are designed to provide robust localization cues under reverberant and multi-source acoustic conditions.
Nevertheless, \textcolor{black}{our contribution is not tied to a particular backend and seamlessly generalizes} to any method capable of producing analogous maps, e.g.,~\ac{MUSIC}~\cite{schmidt1986multiple},~\ac{ESPRIT}~\cite{roy1989esprit} and~\ac{DNN}-based models~\cite{neural-SRP, 9287466}.}\looseness=-1

\subsection{SRP-PHAT}
\label{subsec:srp_phat}
A prominent example is~\ac{SRP-PHAT}~\cite{dibiase2001robust}, which steers the array over a spatial grid of hypothesized source locations.
This process yields a spatial likelihood map, whose peaks correspond to potential source directions. To construct this map, the phase-normalized cross-correlation is evaluated for each microphone pair~$(m, m')$ at every~\ac{T-F} bin\looseness=-1
\begin{equation}
    \Psi_{mm'}(t,f) = \frac{X_m(t,f) X_{m'}^*(t,f)}{|X_m(t,f) X_{m'}^*(t,f)|},
    \label{eq:gcc_phat}
\end{equation}
where $|\!\cdot\!|$ denotes the magnitude of a complex number. The normalization enforces unit magnitude across frequencies, leaving only phase differences. 
Then, given a hypothesized direction~$\mathbf{p}=[\phi, \theta]^{\mathsf{T}}\in\mathbb{R}^2$, the expected~\ac{TDOA} between a pair of microphones~$m$ and~$m'$ is
\begin{equation}
    \label{eq:tdoa}
\tau_{mm'}(\mathbf{p}) = (\mathbf{d}_m - \mathbf{d}_{m'})^{\mathsf{T}}\mathbf{u}(\mathbf{p})/c,
\end{equation}
where~$\mathbf{d}_m\in\mathbb{R}^3$ and~$\mathbf{d}_{m'}\in\mathbb{R}^3$ are the microphone positions,~$\mathbf{u}(\mathbf{p})\!=\![\sin(\phi)\cos(\theta), \sin(\phi)\sin(\theta), \cos(\phi)]^{\mathsf{T}}\in\mathbb{R}^3$ is the unit vector pointing toward the source and~$c$ is the speed of sound. 
The SRP likelihood map at direction $\mathbf{p}$ is computed using~$Q=M(M-1)/2$ non-redundant microphone pairs\looseness=-1
\begin{align}
   \label{eq:likelihood_map}
&\Phi_{\mathrm{SRP-PHAT}}(\mathbf{p},t) = \\
&\frac{1}{QF}
   \sum_{m'<m}
    \sum_{f=1}^F
   \Re \Big\{ \Psi_{mm'}(t,f) 
   e^{-j\omega_f \tau_{mm'}(\mathbf{p})} \Big\}. \nonumber
\end{align}
where $\Re\{\cdot\}$ denotes the real part, and $\omega_f = 2\pi f$. Multiplying by~$e^{-j\omega_f \tau_{mm'}(\mathbf{p})}$ virtually steers the microphone pair $(m,m')$ towards the candidate direction. Averaging across~\ac{T-F} bins and microphone pairs produces a spatial likelihood map $\Phi_{\mathrm{SRP-PHAT}}$ whose peaks indicate the most probable source directions.\looseness=-1

\subsection{SRP-DNN}
\label{sec:srp-dnn}
Classic~\ac{SSL} algorithms often struggle when multiple sources are simultaneously active, as interference can lead to less pronounced spatial peaks, peak merging, or even spurious responses. Reverberation and ambient noise further exacerbate these issues, making accurate source detection and localization increasingly difficult.
To this end, the SRP-DNN~\cite{srp_dnn} uses a deep learning model designed to suppress acoustic interference and estimate the direct-path~\acp{IPD} for each microphone pair $(m,m')$, given by:\looseness=-1
\begin{equation}
\label{eq:IPD}
\mathrm{IPD}_{mm'}(f;\mathbf{p}^*) = e^{j \omega_f \tau_{mm'}(\mathbf{p}^*)}.
\end{equation}
Note that in anechoic, noise-free rooms, the phase-normalized cross-correlation defined in~\eqref{eq:gcc_phat} satisfies ${\Psi}_{mm'}(t, f)\!=\!\mathrm{IPD}_{mm'}(f;\mathbf{p}^*)$ for a true source located at~$\mathbf{p}^*$. Consequently, the likelihood in~\eqref{eq:likelihood_map} is maximized whenever~$\mathbf{p}=\mathbf{p}^*$.\looseness=-1

At its core, SRP-DNN adopts a~\ac{CRNN} to extract \acp{IPD} from noisy and reverberant microphone signals. 
Let $\mathbf{r}_{mm'} (\mathbf{p}^*)\in\mathbb{R}^{2F\times1}$ denote the real-valued vector form of the direct-path~\ac{IPD}
\begin{equation}
\mathbf{r}_{mm'}(\mathbf{p}^*) =
\begin{aligned}[t]
\big[&\cos(\omega_{1} \tau_{mm'}(\mathbf{p}^*)), \sin(\omega_{1} \tau_{mm'}(\mathbf{p}^*)),\ldots,\\
        &\cos(\omega_{F} \tau_{mm'}(\mathbf{p}^*)), \sin(\omega_{F} \tau_{mm'}(\mathbf{p}^*)) \big]^{\mathsf{T}}.
\end{aligned}
\end{equation}
The model is trained to estimate a summed direct-path~\ac{IPD} representation encoding the contributions of all $K$ sources
\begin{equation}
\label{eq:training_target}
    \mathbf{R}_{mm'}(t)=\sum_{k=1}^K\pi_k(t)\mathbf{r}_{mm'}(\mathbf{p}^*_k),
\end{equation}
where $\pi_k(t)\!\in\![0,1]$ denotes the activity probability of the $k$-th source at the $t$-th frame. The SRP-DNN model is optimized by minimizing the~\ac{MSE} between the prediction $\widehat{\mathbf{R}}_{mm'}(t)$ and the training target at~\eqref{eq:training_target}.
Analogously to \eqref{eq:likelihood_map}, the spatial likelihood map is given by
\begin{equation}
\label{eq:srp-dnn-likelihood-map}
\Phi_{\mathrm{SRP\text{-}DNN}}(\mathbf{p},t)\!=\! 
\frac{1}{QF}
\sum_{m'<m}\!
\widehat{\mathbf{R}}^{\mathsf{T}}_{mm'}(t) \mathbf{r}_{mm'}(\mathbf{p}).
\end{equation}
Note that~\eqref{eq:srp-dnn-likelihood-map} becomes equivalent to~\eqref{eq:likelihood_map} by substituting the network estimate $\widehat{\mathbf{R}}_{mm'}(t)$ with
\begin{equation}
 \widehat{\mathbf{R}}^\textrm{SRP}_{mm'}(t)=
\begin{aligned}[t]
       \big[&\Re(\Psi_{mm'}(t,1)),\Im(\Psi_{mm'}(t,1)),\ldots,\\
       &\Re(\Psi_{mm'}(t,F)),\Im(\Psi_{mm'}(t,F))\big]^\mathsf{T}.
\end{aligned}
\end{equation}

\subsection{Source Localization from the Likelihood Map}
We now describe how to extract source position estimates from a given spatial likelihood map (e.g., as defined in~\eqref{eq:likelihood_map} and~\eqref{eq:srp-dnn-likelihood-map}). 
In the single-source case, the estimated position can be simply determined by the peak of the likelihood map
\begin{equation}
\hat{\mathbf{p}}(t) = \underset{\mathbf{p} \in \mathcal{G}}{\operatorname{arg\,max}} \; \Phi(\mathbf{p},t), 
\label{eq:srp_argmax}
\end{equation}
where $\mathcal{G}$ denotes a grid of candidate directions.
In practice, this estimator cannot properly recover multiple overlapping speakers in adverse acoustic conditions.
An alternative approach~\cite{srp_dnn, 7445871, brutti2010multiple, diaz2020source}, is to iteratively detect the dominant source from the likelihood map, suppress its spatial contribution, and then re-estimate the map to uncover new sources. Since the true number of sources is often unknown, a stopping criterion is essential. However, the stopping threshold is typically chosen heuristically, without performance guarantees.\looseness=-1

We adopt the iterative detection–localization procedure~\cite{srp_dnn}, as summarized in Algorithm~\ref{alg:iterative_ssl}, where time index is omitted for brevity.
The algorithm runs for $K$ iterations if the source count is known; otherwise, it stops when a threshold is met.
In Section~\ref{sec:proposed_method}, we show that this threshold can be selected through a principled risk-controlling framework.\looseness=-1

We note that although Algorithm~\ref{alg:iterative_ssl} is presented in the context of SRP-type methods, the~\ac{UQ} frameworks developed in this paper are general. They can be readily adapted, with only minor modifications, to any~\ac{SSL} approach yielding a sequence of detected~\acp{DOA} and their corresponding likelihood maps.
\looseness=-1

\setlength{\abovecaptionskip}{5pt}
\setlength{\belowcaptionskip}{5pt}
\setlength{\textfloatsep}{5pt}
{
\begin{algorithm}[t]
\caption{Iterative Source Detection and Localization}
\label{alg:iterative_ssl}
\begin{algorithmic}[1]
\Require Spatial grid $\mathcal{G}$, likelihood maps $\Phi(\mathbf{p},t)$ for $\mathbf{p}\!\in\!\mathcal{G}$; 
Stopping threshold $\beta_{\mathrm{TH}}$ and maximum count $K_{\textrm{max}}$ or known count $K$; Minimum angular separation between sources $d$, $\widehat{\mathbf{R}}_{mm'}$ is estimated \ac{IPD}; $\gamma$ is an attenuation factor to avoid over suppression.
\Ensure 
Detected~\acp{DOA} $\{\hat{\mathbf{p}}_1, \ldots, \hat{\mathbf{p}}_k\}$ and the corresponding likelihood maps $\{\Phi^{(1)}, \ldots, \Phi^{(k)}\}$
\State Initialize $\Phi^{(1)}(\mathbf{p}) \gets \Phi(\mathbf{p})$
\State Initialize $\mathcal{F}_1\gets\mathcal{G}$
\For{$k \gets 1$ \textbf{to} $K_{\mathrm{max}}$}\Comment{or $K$ loops at known count}
  \State $\hat{\mathbf{p}}_k \gets 
          \operatorname*{arg\,max}_{\mathbf{p}\in\mathcal{F}_k} \Phi^{(k)}(\mathbf{p})$
  \If{$\Phi^{(k)}(\hat{\mathbf{p}}_k) < \beta_{\mathrm{TH}}$} \Comment{only for unknown count}
        \State \textbf{break}
  \EndIf
  \State $\hat{\beta}_k \gets \gamma\cdot\Phi^{(k)}(\hat{\mathbf{p}}_k)$
  \vspace{0.3em}
  \State $\widehat{\mathbf{R}}_{mm'} \leftarrow \widehat{\mathbf{R}}_{mm'} - \hat{\beta}_k \cdot \mathbf{r}_{mm'}(\hat{\mathbf{p}}_k)$
    \State \textsc{Update}\;$\Phi^{(k+1)}$\hfill\Comment{Eq.~\eqref{eq:srp-dnn-likelihood-map}}
\State $\mathcal{F}_{k+1} \! \gets \! \big\{\mathbf{p}\!\in\!\mathcal{G}\!:\! 
         |\phi-\hat{\phi}_{i}| \ge d\wedge|\theta-\hat{\theta}_{i}| \ge d,
         \;\forall i\!<\!k\big\}$                 
\EndFor
\State \Return $\{\hat{\mathbf{p}}_1, \ldots, \hat{\mathbf{p}}_k\}, 
\,\,\{\Phi^{(1)}, \ldots, \Phi^{(k)}\}$
\end{algorithmic}
\end{algorithm}
}

\subsection{Uncertainty Quantification}
A key limitation of many~\ac{SSL} methods is that they yield only point estimates, offering no indication of confidence. Incorporating~\ac{UQ} is therefore highly beneficial, as it enables more informed downstream decisions by quantifying the confidence in each localization result. 
In the context of~\ac{SSL}, uncertainty can be naturally expressed using prediction regions constructed directly on top of the spatial likelihood map. In particular, a more compact region indicates higher confidence in the corresponding localization estimate. 
\textcolor{black}{Note that~\ac{CP} provides marginal statistical guarantees on average over the entire data population, and does not bound the error for specific samples. In the experimental part, we demonstrate the correlation between the predicted region and the localization error, and show that larger areas are obtained for later detected sources and higher reverberation and noise, as expected.}

When the number of sources is unknown, we would like to account also for the uncertainty in detecting the correct number of sources.
To enable statistical guarantees for prediction coverage and reliable control over missed detections, our framework relies on~\ac{CP} and its extensions. The following section provides a detailed background on these methods.\looseness=-1

\section{Background On Conformal Prediction}
\label{sec:conformal_prediction}
\subsection{Overview}
In recent years, \ac{CP}~\cite{gammerman1998transduction, vovk2005algorithmic, angelopoulos2024theoretical} 
has gained popularity as it provides a distribution-free framework for constructing predictive intervals with finite-sample guarantees, independent  of the underlying model or data distribution. Its appeal stems from being a generic post-hoc approach that requires only data exchangeability, i.e., the joint distribution of data samples is invariant under any permutation. This generality has enabled successful applications primarily in computer vision~\cite{angelopoulos2020uncertainty, silva2025conformal} and natural language processing~\cite{fisch2020efficient, quach2024conformal}, and more recently in speech processing~\cite{CP_MMGP, khurjekar2023uncertainty, khurjekar2024multi, gnn_ssl}.

Split~\ac{CP}~\cite{vovk2005algorithmic,papadopoulos2002inductive} is a practical variant of~\ac{CP} that employs a simple data-splitting scheme. The available samples are divided into a training set, used to fit a predictive model~$\hat f$, and a separate calibration set
$\mathcal{D}_{\mathrm{cal}} = \{(X_i, Y_i)\}_{i=1}^n,(X_i, Y_i)\sim \mathcal{P}$, where $\mathcal{P}$ is an arbitrary distribution on
$\mathcal{X}\times\mathcal{Y}$.
In this context, \(X_i \in \mathcal{X}\) denotes a feature vector and \(Y_i \in \mathcal{Y}\) the corresponding response.
Given a fresh test pair \((X_{n+1}, Y_{n+1}) \sim \mathcal{P}\), the goal is to construct a statistically valid~\ac{PI}, i.e., \(\mathcal{C}(X_{n+1})\), that contains the unobserved response \(Y_{n+1}\) with high probability. To this end, a nonconformity score function $s(x, y) \in \mathbb{R}$ is used to measure the agreement between $y$ and $x$ based on a trained model $\hat{f}$. Better consistency leads to a smaller score. The choice of the nonconformity score is task-dependent, but a common choice for regression problems is the residual error function $s(x, y) = \big| y - \hat{f}(x) \big|$. 
Let \(s_i = s(X_i, Y_i)\) denote the score associated with the $i$-th calibration sample, the prediction interval for an unseen feature vector \(X_{n+1}\) is 
\begin{equation}
    \mathcal{C}(X_{n+1}) =  [\hat{f}(X_{n+1})-\hat{q}_{1-\alpha},\; \hat{f}(X_{n+1})+\hat{q}_{1-\alpha}],
    \label{eq:prediction_interval}
\end{equation}
where \(\hat{q}_{1-\alpha}\) is the \((1-\alpha)(1 + 1/n)\)-quantile of the calibration scores \(\{s_1, \ldots, s_n\}\).
Under the sole assumption that the calibration and test samples $\{(X_i, Y_i)\}_{i=1}^{n+1}$ are exchangeable,
split~\ac{CP}~\cite{vovk2005algorithmic} provides a finite-sample coverage
guarantee:
\begin{equation}
    \label{eq:conforal_prediction}
    \mathbb{P}(Y_{n+1}\in\mathcal{C}(X_{n+1})) \ge 1-\alpha.
\end{equation}
By the law of total expectation, this is equivalent to $\mathbb{E}[\mathbbm{1}\{Y_{n+1}\notin\mathcal{C}(X_{n+1}\}]\le\alpha$
with an expectation over the \ac{MC} loss $\mathbbm{1}\{Y_{n+1}\notin\mathcal{C}(X_{n+1})\}$.\looseness=-1

\subsection{Conformal Risk Control}
\label{subsec:crc}
While~\ac{CP} controls the~\ac{MC} loss, it may not be the most natural choice for many tasks. \ac{CRC}~\cite{crc} extends~\ac{CP} to arbitrary bounded loss functions
$L(\lambda)$ that are non-increasing in a tuning parameter $\lambda\!\in\!\Lambda\!\subset\!\mathbb{R}\cup\{\pm\infty\}$, meaning $L(\lambda_2) \le L(\lambda_1)$ for $\lambda_1 \!<\! \lambda_2$.
Subsequently, the associated risk function is defined as\looseness=-1
\begin{equation}
    \mathcal{R}(\lambda) = \mathbb{E}[L(\lambda)], \quad\lambda\!\in\!\Lambda,
    \label{eq:risk_def}
\end{equation}
assuming the existence of $\lambda_{\min}\!\in\! \Lambda$ such that $\mathcal{R}(\lambda_{\min})\!=\!0$.

Let the random variables $L_i:\Lambda \to (-\infty,B]$ for $i=1,\ldots,n+1$ be defined by~$L_i({\lambda})=L(X_i,Y_i;\lambda)$ and define the corresponding empirical risk over the calibration data as 
\begin{equation}
\label{eq:empirical_risk}
\widehat{\mathcal{R}}_n({\lambda})=\frac{1}{n}\sum_{i=1}^{n} L_i({\lambda}).
\end{equation}
Given a target risk tolerance $\alpha\in(-\infty,B)$, the~\ac{CRC} procedure selects the threshold
\begin{equation}
    \label{eq:optimal_lambda_hat_CRC}
    \hat{\lambda}=\inf \Big\{ \lambda:\widehat{\mathcal{R}}_n({\lambda})\le \alpha-\tfrac{B-\alpha}{n}\Big\},
\end{equation}
which controls the risk on unseen data as stated in the theorem below.\looseness=-1
\begin{theorem}[Conformal Risk Control~\cite{crc}]
\label{theo:risk_control}
    Suppose $L_i(\lambda)$ is non-increasing in $\lambda$, right-continuous, satisfying 
    $L_i(\lambda_{\max})\le\alpha$ for $\lambda_{\mathrm{max}}=\sup \Lambda\in\Lambda$ and $\sup_\lambda L_i(\lambda)\le B<\infty$ almost surely. 
    Then $\mathbb{E}[L_{n+1}(\hat{\lambda})]\le\alpha$.
\end{theorem}

\subsection{Advanced Risk-Control Methods}
\label{subsec:advanced_crc}
\ac{CRC} controls monotonic risks at level $\alpha$, but a stronger version would provide high-probability risk control\looseness=-1
\begin{equation}
    \mathbb{P}\!\left( \mathcal{R}({\lambda})\le \alpha \right)\ge 1-\delta.
    \label{eq:risk_control_property}
\end{equation}
meaning that with at least $1-\delta$ probability, any drawn calibration set $\mathcal{D}_{\mathrm{cal}}$ yields an expected loss that does not exceed the risk limit $\alpha$.
Indeed, \ac{RCPS}~\cite{bates2021distribution,einbinder2025semi} offers a rigorous method for constructing prediction functions that achieves the property at~\eqref{eq:risk_control_property}.
\ac{LTT}~\cite{angelopoulos2025learn} extends~\ac{RCPS} by removing the monotonicity requirement on the risk and enabling the calibration of multi-dimensional threshold configurations, $\boldsymbol{\lambda}=(\lambda_1,\ldots,\lambda_D)$ defined over a continuous space 
$\boldsymbol{\Lambda}=\Lambda_1\times\dots\times\Lambda_D$. This allows simultaneous control of multiple risks, $\mathcal{R}_{\ell}\le\alpha_{\ell}$ for $\ell=1,\ldots,N_c$. The key idea is to cast the configuration selection as a~\ac{MHT} procedure, testing a finite set of configurations $\boldsymbol{\Lambda}_g\subseteq\boldsymbol{\Lambda}$.
For each configuration~$\boldsymbol{\lambda}_j \in \boldsymbol{\Lambda}_g$ 
consider the null hypothesis\looseness=-1
\begin{equation}
    \mathcal{H}_j : \exists \ell \in \{1, \ldots, N_c\} \;\; \text{such that} \;\;
    \mathcal{R}_{\ell}({\boldsymbol{\lambda}_j}) > \alpha_{\ell}.
    \label{eq:null_hypothesis}
\end{equation}
By construction, \emph{rejecting} $\mathcal{H}_{j}$ certifies that, for configuration $\boldsymbol{\lambda}_j$, all risks are controlled, i.e., $\mathcal{R}_{\ell}({\boldsymbol{\lambda}_j})\leq \alpha_{\ell}$ for every $\ell$.
Let $\boldsymbol{\Lambda}_{\mathrm{rej}} \subseteq \boldsymbol{\Lambda}_g$ denote the true rejection set. The~\Ac{LTT} framework provides a principled method for determining an estimated rejection set $\hat{\boldsymbol{\Lambda}}_{\mathrm{rej}} \subseteq \boldsymbol{\Lambda}_g$ based on calibration data. Importantly, it controls the~\ac{FWER} at level $\delta$, meaning that the probability of making at least one false rejection among all tested hypotheses is bounded by $\delta$.

For constructing the rejection set, we first need to derive a finite-sample valid $p$-value~$p$, i.e., $\mathbb{P}(p\!\le\! u)\!\le\! u$ for $u\in[0,1]$ under $\mathcal{H}_j$ (see Appendix~\ref{appendix:p_values} for further details). For a given configuration $\boldsymbol{\lambda}_j$ and risk tolerance $\alpha_{\ell}$, let~$p(\{L^{\ell}_{i}(\boldsymbol{\lambda}_j)\}_{i\in\mathcal{D}_{\mathrm{cal}}}; \:\alpha_{\ell})$ denote the $p$-value of the $j$-th null hypothesis 
$\mathcal{H}_{j,\ell}:\mathcal{R}_{\ell}({\boldsymbol{\lambda}_j}) > \alpha_{\ell}$ associated with the $l$-th loss. 
The $p$-value of the aggregated null hypothesis 
in~\eqref{eq:null_hypothesis} can be computed as~\cite{angelopoulos2025learn}\looseness=-1
\begin{equation}
    p^{\mathrm{cal}}_j
    = \max_{{\ell}=1,\ldots,N_c} \;
      p(\{L^{\ell}_{i}(\boldsymbol{\lambda}_j)\}_{i\in\mathcal{D}_\textrm{cal}}; \:\alpha_{\ell}), 
      \quad \boldsymbol{\lambda}_j \in \boldsymbol{\Lambda}_g.
      \label{eq:p_value}
\end{equation}
Then, whenever $p_j^{\mathrm{cal}} \!<\! \delta$, we declare that all risks $\mathcal{R}_{\ell}({\boldsymbol{\lambda}_j}),\; \ell=1,\ldots,N_c$, are simultaneously controlled.

Given valid \(p\)-values, a rejection set 
\(\hat{\boldsymbol{\Lambda}}_{\mathrm{rej}}\) is constructed by applying an~\ac{FWER}-controlling algorithm \(\mathcal{A} \!:\! [0,1]^{|\boldsymbol{\Lambda}_g|} \!\to\! 2^{\boldsymbol{\Lambda}_g}\). In essence, the algorithm $\mathcal{A}$ takes as input the collection of $p$-values and returns a set of rejected configurations
$\hat{\boldsymbol{\Lambda}}_{\mathrm{rej}} = \mathcal{A}(p^{\mathrm{cal}}_1,\ldots,p^{\mathrm{cal}}_{|\boldsymbol{\Lambda}_g|})$,
while ensuring that the~\ac{FWER} does not exceed \(\delta\). Formally,~\ac{LTT} has the following guarantees.
\begin{theorem}
\label{theo:ltt_theorem}
Suppose $\mathbb{P}(p^{\mathrm{cal}}_j\le u)\le u,\; u\in[0, 1]$ for all $j$ under $\mathcal{H}_{j}$. Let $\mathcal{A}$ be an~\ac{FWER}-controlling algorithm at level $\delta$. Then $\hat{\boldsymbol{\Lambda}}_{\mathrm{rej}} = \mathcal{A}(p_1^{\mathrm{cal}},\ldots,p_{|\boldsymbol{\Lambda}_g}|^{\mathrm{cal}})$, satisfies the following
\begin{equation}
    \mathbb{P}\!\left(
    \sup_{\hat{\boldsymbol{\lambda}} \in \hat{\boldsymbol{\Lambda}}_{\mathrm{rej}}}
    \{ \mathcal{R}_{\ell}(\hat{\boldsymbol{\lambda}}) \} \leq {\alpha_{\ell}}\right) \;\geq\; 1 - \delta,\;\; {\ell}=1,\ldots,N_c.
\label{eq:alpha_delta_fwer}
\end{equation}
\label{theo:FWER_algo}
\end{theorem}
\vspace{-2em}
\subsection{Efficient Risk Control via Pareto-Testing}
\label{subsec:pareto_testing}
The~\ac{LTT} framework relies on the choice of~\ac{FWER}-controlling algorithm. A simple option is the Bonferroni correction, $\mathcal{A}(p^{\mathrm{cal}}_1, \ldots,p^{\mathrm{cal}}_{|\boldsymbol{\Lambda}_g|})
= \{\boldsymbol{\lambda}_j\in\boldsymbol{\Lambda}_g: p^{\mathrm{cal}}_j<\delta/|\boldsymbol{\Lambda}_g|\},$
which becomes increasingly conservative as~$|\boldsymbol{\Lambda}_g|$ grows, thereby reducing its statistical power, i.e., the ability to reject when the null hypothesis is false. 
Advanced testing methods exploit prior knowledge of the  structure of the configuration space to identify promising configurations that are likely to control the risks. For example,~\ac{FST}~\cite{angelopoulos2025learn} orders configurations, from the most to the least promising and tests them sequentially until a null hypothesis fails to be rejected. However, the statistical power of methods such as~\ac{FST} still depends critically on correctly identifying promising configurations and determining an effective testing order, tasks that become challenging when the configuration space $|\boldsymbol{\Lambda}_g|$ is large and possibly lacking structure.
The difficulty intensifies further in settings that involve both constrained objectives to be controlled and unconstrained objectives to be optimized.\looseness=-1 

Building on the \ac{LTT} framework, Pareto-Testing~\cite{DBLP:conf/iclr/Laufer-Goldshtein23} delivers a computationally and statistically efficient \ac{FWER}-controlling algorithm that satisfies~\eqref{eq:alpha_delta_fwer} and simultaneously achieves strong performance on multiple additional objective functions.
The core idea is to partition the calibration set into two disjoint subsets,
$\mathcal{D}_{\mathrm{cal}}=\mathcal{D}_{\mathrm{opt}} \cup\mathcal{D}_{\mathrm{tst}}$, where $\mathcal{D}_{\mathrm{opt}}$ is used to define an ordered sequence of candidate configurations, and $\mathcal{D}_{\mathrm{tst}}$ is used to perform hypothesis testing. Assume \(N\) objective functions $\{\mathcal{R}_1, \ldots, \mathcal{R}_N\}$, where the first \(N_c\) are constrained and the remaining \(N-N_c\) are unconstrained. The set $\mathcal{D}_{\mathrm{opt}}$ is employed to construct the Pareto frontier with respect 
to all objective functions $\{\mathcal{R}_1, \ldots, \mathcal{R}_N\}$. This frontier, denoted by $\boldsymbol{\Lambda}_{\mathrm{par}}$, consists of configurations that are not \emph{dominated} by any other configuration. Formally, for any $\boldsymbol{\lambda},\boldsymbol{\lambda}' \in \boldsymbol{\Lambda}_g$,
we say that $\boldsymbol{\lambda}'$ dominates $\boldsymbol{\lambda}$
(denoted as $\boldsymbol{\lambda}' \prec \boldsymbol{\lambda}$) if
$\mathcal{R}_{\ell}({\boldsymbol{\lambda}'}) \le
\mathcal{R}_{\ell}({\boldsymbol{\lambda}})$ for all $\ell$ and $\mathcal{R}_{\ell}({\boldsymbol{\lambda}'}) < \mathcal{R}_{\ell}({\boldsymbol{\lambda}})$ for at least one $\ell$.
The Pareto frontier is defined as
\begin{equation}
\label{eq:lambda_pareto}
    \boldsymbol{\Lambda}_{\mathrm{par}}
= \Bigl\{
\boldsymbol{\lambda}\in\boldsymbol{\Lambda} :
\{\boldsymbol{\lambda}'\in\boldsymbol{\Lambda} :
\boldsymbol{\lambda}' \prec \boldsymbol{\lambda},\;
\boldsymbol{\lambda}' \neq \boldsymbol{\lambda}\}
= \emptyset
\Bigr\}.
\end{equation}
In practice, the Pareto frontier may be obtained using standard Pareto optimization methods, including brute-force grid search (see Algorithm~\ref{alg:recover_pareto} in Appendix~\ref{appendix:pareto_opt_set}) or multi-objective Bayesian optimization algorithms~\cite{lindauer2022smac3}.
The configurations in $\boldsymbol{\Lambda}_{\mathrm{par}}$ are ranked in ascending order according to their estimated $p$-values computed on $\mathcal{D}_{\mathrm{opt}}$, after which \ac{FST} is applied to the ordered set using $\mathcal{D}_{\mathrm{tst}}$. As long as $p_j^{\mathrm{tst}} < \delta$, the hypothesis $\mathcal{H}_j$ is rejected, and the corresponding configuration $\boldsymbol{\lambda}_j$ is added to $\hat{\boldsymbol{\Lambda}}_{\mathrm{rej}}$. This procedure halts at the configuration for which the null hypothesis cannot be rejected. In a final step, a refined set of admissible configurations is obtained by filtering the Pareto-optimal configurations over $\hat{\boldsymbol{\Lambda}}_{\mathrm{rej}}$, this time considering only the free objective functions $\{\mathcal{R}_{N_c+1}, \ldots, \mathcal{R}_N\}$. The resulting subset, denoted by $\boldsymbol{\Lambda}^*$, contains all validated configurations that are Pareto-optimal with respect to the free objective functions. 
\looseness=-1

\section{Proposed Approach}
\label{sec:proposed_method}
This section introduces our complementary methods, which address the multiple-source localization problem of Section~\ref{sec:problem_formulation}.
Both methods combine generic spatial likelihood maps with risk-controlling selection methods to provide principled~\ac{UQ}.
The first method assumes a known number of sources and uses \ac{CRC} (Section~\ref{subsec:crc}) to construct tight prediction regions around each estimated position, which quantify localization uncertainty under a user-specified coverage level.
The second method handles the unknown-source count setting by jointly performing source detection and localization using
Pareto-Testing (Section~\ref{subsec:pareto_testing}).
In this case, the objectives extend to accurately estimating the true number of active sources and constructing their corresponding prediction regions, while minimizing false detections and the total prediction area. 
\looseness=-1

\begin{figure*}[t]
    \centering
    \includegraphics[width=\linewidth]{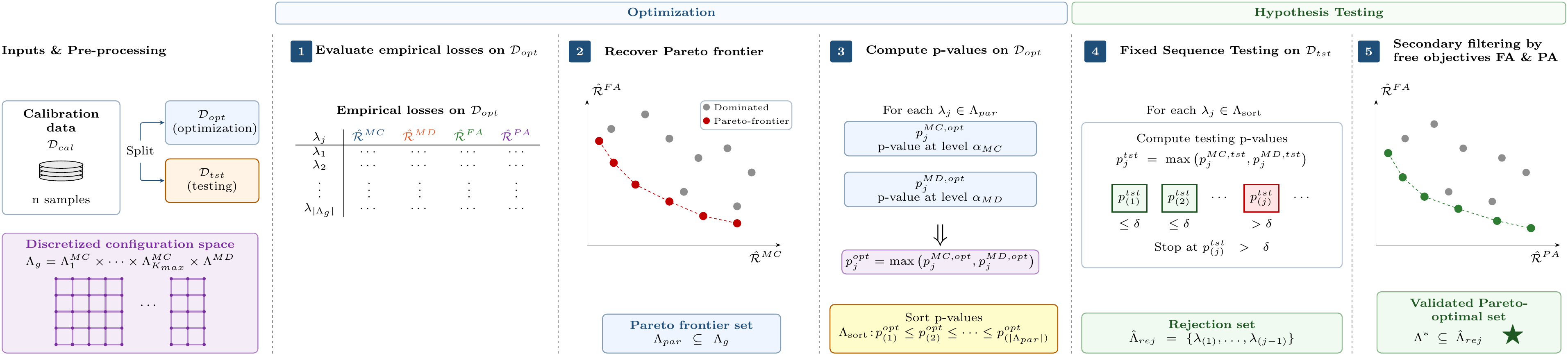}
    \caption{\textcolor{black}{Overview of the PT-SSL-U framework. For ease of visualization, Stage~2 illustrates the procedure using two objectives.}}
    \label{fig:PT-SSL-U-flow}
    \vspace{-0.5cm}
\end{figure*}

\begin{figure}[t]
    \centering
    \includegraphics[width=\linewidth]{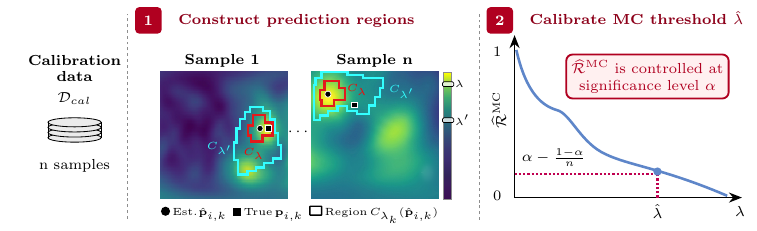}
    \caption{\textcolor{black}{Overview of the CRC-SSL-N framework.}}
    \label{fig:CRC-SSL-N-flow}
    \vspace{-0.5cm}
\end{figure}

\begin{figure}[t]
    \centering
    \includegraphics[width=\linewidth]{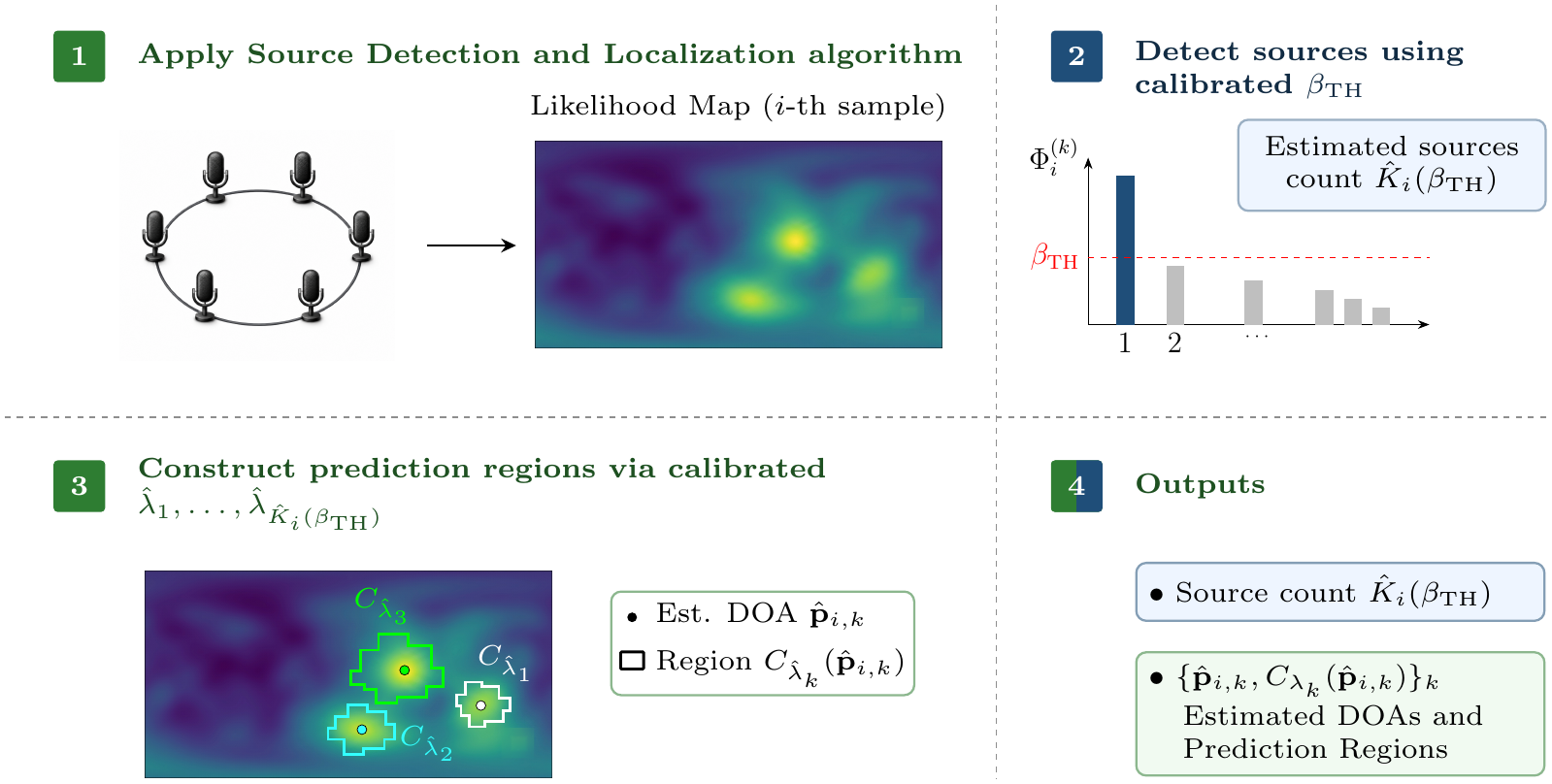}
    \caption{\textcolor{black}{Inference workflow of the CRC-SSL-N and PT-SSL-U frameworks. Components highlighted in blue correspond to the additional steps used by PT-SSL-U.}}

    \label{fig:inference-flow}
\end{figure}

\subsection{Known Source Count}
\label{subsec:known_source_count}
Let~$\{\mathbf{p}^*_{i,1}, \ldots, \mathbf{p}^*_{i,K}\}$ denote the (unordered) set of true source~\acp{DOA} for sample~$i$.  
Assume that a set of $K$ predicted directions~$\{\hat{\mathbf{p}}_{i,1}, \ldots, \hat{\mathbf{p}}_{i,K}\}$,
together with their spatial likelihood maps
$\{\Phi_i^{(1)}, \ldots, \Phi_i^{(K)}\}$, is available. We employ a greedy sequential matching procedure that assigns each predicted position, in order, to its nearest unmatched true position, yielding the pairs
\((\hat{\mathbf{p}}_{i,1}, \mathbf{p}^*_{i,1}),\ldots,(\hat{\mathbf{p}}_{i,K}, \mathbf{p}^*_{i,K})\). For notational simplicity indices are kept unchanged, though the~$k$-th prediction is not necessarily matched to the~$k$-th source. Note that alternative matching mechanisms may also be used, without affecting the risk-controlling framework proposed in the sequel.\looseness=-1

For the~\(k\)-th detected source in sample~\(i\), a prediction region~\(\mathcal{C}_{\lambda_k}(\hat{\mathbf{p}}_{i,k})\) is constructed using a threshold~\(\lambda_k \in \Lambda_k^{\mathrm{MC}}\). 
Each prediction region is defined as a spatially connected region of grid points \(\mathbf{p} \!\in\! \mathcal{G}\) surrounding \(\hat{\mathbf{p}}_{i,k}\). 
Starting from \(\hat{\mathbf{p}}_{i,k}\), the set is iteratively expanded by including any grid point that is adjacent to the current set and satisfies~\(\Phi_i^{(k)}(\mathbf{p}) \ge \lambda_k\). Algorithm~\ref{algo:prediction_set} summarizes this construction.\looseness=-1

Importantly, this formulation differs from the standard construction of~\ac{CP} sets or intervals. Using classification-based ~\ac{CP}, each azimuth–elevation pair on the likelihood map can be treated as a class and included in the prediction set
whenever its likelihood exceeds a threshold~$\lambda_k$. However, the resulting prediction region may be non-contiguous due to spurious peaks arising from noisy and echoic environments. Moreover, in multi-source scenarios, standard~\ac{CP} yields only a single joint region, making it unclear how to split the region into individual sources.
Alternatively, using a regression formulation, a prediction region could be defined as a fixed 2D geometric shape centered at~$\hat{\mathbf{p}}_k$ (e.g., a circle or a square). However, this would impose an unrealistic assumption that the likelihood landscape follows a particular geometric structure.\looseness=-1
 
An~\ac{MC} occurs whenever the true DOA~\(\mathbf{p}^*_{i,k}\) is not included in the associated prediction region. Formally, for the~\(k\)-th source, define the exchangeable random variables 
\(L_{i,k}^{\mathrm{MC}} : \Lambda_k^{\mathrm{MC}} \to \{0,1\}\) for 
\(i = 1,\ldots,n+1\),
\begin{equation}
\label{eq:misc_cov_loss}
    L_{i,k}^{\mathrm{MC}}(\lambda_k)
    = \mathbbm{1}\!\left\{
        \mathbf{p}^*_{i,k} \notin 
        \mathcal{C}_{\lambda_k}(\hat{\mathbf{p}}_{i,k})
      \right\}.
\end{equation}
The corresponding empirical \ac{MC} risk is
\begin{equation}
\label{eq:mis_cov_risk}
    \widehat{\mathcal{R}}_{n,k}^{\mathrm{MC}}(\lambda_k)
    = \frac{1}{n}
      \sum_{i=1}^{n} 
      L_{i,k}^{\mathrm{MC}}(\lambda_k),
    \qquad k=1,\ldots,K.
\end{equation}
For each source $k$, we aim to construct $\mathcal{C}_{\hat{\lambda}_k}$
that guarantees the~\ac{MC} risk on unseen data is at most $\alpha$, namely
\begin{equation}
    \mathbb{E}\bigl[ L^{\mathrm{MC}}_{n+1,k}(\hat{\lambda}_k) \bigr]
    \le \alpha, \qquad k=1,\ldots,K.
\label{eq:mc_risk_control_known_speakers_amount}
\end{equation}
In this setting, smaller thresholds produce larger prediction regions: for any $\lambda_1<\lambda_2$, $\mathcal{C}_{\lambda_2}\subseteq\mathcal{C}_{\lambda_1}$, implying $L(\lambda_1)\le L(\lambda_2)$ since the \ac{MC} loss can only decrease as regions expand. Therefore, $L(\lambda)$ is monotone \emph{non-decreasing} in $\lambda$ and bounded by 1, satisfying the requirements of the \ac{CRC} Theorem~\ref{theo:risk_control}, up to reversed monotonicity relation. Accordingly, we obtain the following~\ac{CRC} selection rule for $\hat{\lambda}_k$ (in reverse direction relative to~\eqref{eq:optimal_lambda_hat_CRC}):\looseness=-1
\begin{equation}
\label{eq:crc_lambda_hat}
\hat{\lambda}_k = \sup \Bigl\{ \lambda_k \in \Lambda^{\mathrm{MC}}_k : \widehat{\mathcal{R}}_{n,k}^{\mathrm{MC}}(\lambda_k) \le \alpha - \tfrac{1-\alpha}{n} \Bigr\}.
\end{equation}

\setlength{\textfloatsep}{5pt}
{
\begin{algorithm}[!t]
\caption{Construction of the prediction regions $\mathcal{C}_{\lambda}(\hat{\mathbf{p}})$}
\label{algo:prediction_set}
\begin{algorithmic}[1]

\Require Predicted~\ac{DOA} $\hat{\mathbf{p}}$; likelihood map $\Phi(\mathbf{p})$; threshold $\lambda$
\Ensure prediction region $\mathcal{C}_{\lambda}(\hat{\mathbf{p}})$

\State Initialize $\mathcal{C}_{\lambda}(\hat{\mathbf{p}}) \gets \{\hat{\mathbf{p}}\}$
\State Initialize a queue $Q \gets [\hat{\mathbf{p}}]$
\While{$Q$ is not empty}
    \State Pop a point $\mathbf{p}'$ from $Q$
    \For{each grid point $\mathbf{p} \in \mathcal{G}$ adjacent to $\mathbf{p}'$}
        \If{$\mathbf{p} \notin \mathcal{C}_{\lambda}(\hat{\mathbf{p}})$ \textbf{ and } {$\Phi(\mathbf{p}) \ge \lambda$}}
            \State Add $\mathbf{p}$ to $\mathcal{C}_{\lambda}(\hat{\mathbf{p}})$
            \State Push $\mathbf{p}$ to $Q$
        \EndIf
    \EndFor
\EndWhile
\State \Return $\mathcal{C}_{\lambda}(\hat{\mathbf{p}})$
\end{algorithmic}
\end{algorithm}
}
This establishes our \textbf{`CRC-SSL-N'} framework, denoting~\ac{CRC} for~\ac{SSL} in the known-source-count setting. Fig.~\ref{fig:CRC-SSL-N-flow} and Fig.~\ref{fig:inference-flow} illustrate the calibration and inference procedures, respectively.
Finally, we define the (relative) \ac{PA} incurred by source $k\!\in\!\{1,\ldots,K\}$ for samples  \(i=1,\ldots,n+1\) as\looseness=-1
\begin{equation}
    L_{i,k}^{\mathrm{PA}}(\lambda_k)
    = \frac{\bigl| \mathcal{C}_{\lambda_k}(\hat{\mathbf{p}}_{i,k}) \bigr|}{|\mathcal{G}|},
\end{equation}
where $|\mathcal{C}_{\lambda_k}(\hat{\mathbf{p}}_{i,k})|$ denotes the area of the prediction region on the likelihood map, whose total spatial support is $|\mathcal{G}|$.
The overall empirical~\ac{PA} is given by
\begin{equation}
\label{eq:empirical_area_risk}
\widehat{{\mathcal{R}}}^{\mathrm{PA}}_{n,k}(\lambda_k)=\frac{1}{n}\sum_{i=1}^nL_{i,k}^{\mathrm{PA}}(\lambda_k),
\end{equation}
where small areas correspond to confident predictions, and vice versa, thereby directly quantifying predictive uncertainty.\looseness=-1  

\subsection{Unknown Source Count}
Assuming that the true number of sources~\(K\) is unknown but satisfies 
\(1 \!\le\! K \le K_{\mathrm{max}}\), 
we denote by $\{\mathbf{p}^*_{i,1}, \ldots, \mathbf{p}^*_{i,K}\}$ the (unordered) true source~\acp{DOA} of the $i$-th sample. Following Algorithm~\ref{alg:iterative_ssl}, for each sample and every threshold  $\beta\in\Lambda^{\mathrm{MD}}$, the number of estimated sources is\looseness=-1 
\begin{equation}
\label{eq:Khat_stopping}
\widehat{K}_{i}(\beta)=\max\!\left\{
k \;:\; \Phi_i^{(j)}\!\left(\hat{\mathbf{p}}_{i,j}\right) \ge \beta,\;\forall j\le k\right\},
\end{equation}
with the convention that $\widehat{K}_{i}(\beta)\!=\!0$ if the set is empty. 

In this setting, we introduce additional loss functions aimed at reliably identifying the exact number of sources, depending on the threshold $\beta$. We define the~\ac{MD} loss that quantifies the number of undetected sources. Let $\boldsymbol{\Lambda}=
\Lambda_1^{\mathrm{MC}}
\times \cdots \times
\Lambda_{K_{\mathrm{max}}}^{\mathrm{MC}}
\times
\Lambda^{\mathrm{MD}}$ denote an augmented configuration space, and define the sequence of random variables $L_i^{\mathrm{MD}}\!:\!\boldsymbol{\Lambda}\!\to\!\{0, \ldots,K\}$
for $i=1,\ldots,n+1$, counting the number of missed sources among the $K$ true sources\looseness=-1
\begin{equation}
    \label{eq:md_loss}
    L_{i}^{\mathrm{MD}}(\boldsymbol{\lambda})=\max\!\left\{0, K-\widehat{K}_i(\beta) \right\}
\end{equation}

In addition, we define the~\ac{FA} loss as the sequence of random variables 
$L_i^{\mathrm{FA}}\!:\!\boldsymbol{\Lambda}\!\to\!\{0,\ldots,K_{\mathrm{max}}-K\}$ for $i=1,\ldots,n+1$
counting the number of spurious peaks
among the detected sources
\begin{equation}
\label{eq:explicit_fa_loss}
L_{i}^{\mathrm{FA}}(\boldsymbol{\lambda})
= \max\!\left\{0, \widehat{K}_i(\beta)-K\right\}.
\end{equation}

Accordingly, we refine the definition of the~\ac{MC} loss since measuring coverage is meaningful only for true detected sources. Thus, we consider only predicted peaks whose likelihood exceeds the detection threshold $\beta$. For a given threshold configuration $\boldsymbol{\lambda}$, the~\ac{MC} loss for the $k$-th source is defined as\looseness=-1
\begin{equation}
L^{\mathrm{MC}}_{i,k}({\boldsymbol{\lambda}})
= \mathbbm{1}\!\left\{
\mathbf{p}^*_{i,k} \notin
\mathcal{C}_{\lambda_k}\bigl(\hat{\mathbf{p}}_{i,k}\bigr)
\;\land\;
\Phi_i^{(k)}(\hat{\mathbf{p}}_{i,k}) \ge \beta
\right\},
\end{equation}
and the overall~\ac{MC} loss is obtained by summing over all true detected sources\looseness=-1
\begin{equation}
L^{\mathrm{MC}}_{i}\bigl({\boldsymbol{\lambda}}\bigr)
    = \sum_{k=1}^{{\min\{K, \widehat{K}_i(\beta)}\}}
      L^{\mathrm{MC}}_{i,k}\bigl({\boldsymbol{\lambda}}\bigr).
\label{eq:mis_cov_loss_averaged}
\end{equation}
The \ac{PA} loss of the $i$-th sample is defined as the average area over all detected source\looseness=-1
\begin{equation}
\label{eq:area_loss}
L^{\mathrm{PA}}_{i}\bigl({\boldsymbol{\lambda}}\bigr)
    = \frac{1}{\widehat{K}_i(\beta)}\sum_{k=1}^{\widehat{K}_i(\beta)}
      \frac{1}{|\mathcal{G}|}
\bigl|\mathcal{C}_{\lambda_k}\bigl(\hat{\mathbf{p}}_{i,k}\bigr) \bigr|.
\end{equation}
For $\widehat{K}_i(\beta)\!=\!0$, the loss is set to $L^{\mathrm{PA}}_{i}(\boldsymbol{\lambda})\!=\!0$ by convention.

With loss functions in place, 
the unknown-source count setting is cast into the following constrained~\ac{MOO} problem:
\begin{equation}
\label{eq:optimization_problem_final}
\begin{aligned}
\hat{\boldsymbol{\lambda}}
&= \mathop{\arg\min}_{\boldsymbol{\lambda} \in \boldsymbol{\Lambda}}
   \bigl(
      \mathcal{R}^{\mathrm{FA}}({\boldsymbol{\lambda}}),\,
      \mathcal{R}^{\mathrm{PA}}({\boldsymbol{\lambda}})
   \bigr) \\
&\text{s.t.}\;
   \mathbb{P}\!\left(
     \mathcal{R}^{\mathrm{MD}}({\boldsymbol{\lambda}})
     \le \alpha_{\mathrm{MD}}
   \right) \ge 1-\delta, \\
&\hphantom{\text{s.t.}\;}
   \mathbb{P}\!\left(
     \mathcal{R}^{\mathrm{MC}}({\boldsymbol{\lambda}})
     \le \alpha_{\mathrm{MC}}
   \right) \ge 1-\delta.
\end{aligned}
\end{equation}
Namely, the goal is to identify configurations $\hat{\boldsymbol{\lambda}}\!\in\!\boldsymbol{\Lambda}$ such that, with high probability~$1\!-\!\delta$,
the~\ac{MD} and~\ac{MC} risks on unseen data are controlled while minimizing the~\ac{PA} and~\ac{FA} losses.\looseness=-1

Toward this end, Pareto-Testing procedure is employed. 
Given a calibration set \(\mathcal{D}_{\mathrm{cal}}\), it is partitioned into disjoint subsets \(\mathcal{D}_{\mathrm{cal}} \!=\! \mathcal{D}_{\mathrm{opt}} \cup \mathcal{D}_{\mathrm{tst}}\). 
Using the optimization subset $\mathcal{D}_{\mathrm{opt}}$, all four objective functions (both constrained and unconstrained) are evaluated over a discretized configuration space $\boldsymbol{\Lambda}_g\subseteq\boldsymbol{\Lambda}$.
We recover the Pareto frontier~\eqref{eq:lambda_pareto} using Algorithm~\ref{alg:recover_pareto} (Appendix~\ref{appendix:pareto_opt_set}), yielding the set~$\boldsymbol{\Lambda}_{\mathrm{par}}\!\subseteq\!\boldsymbol{\Lambda}_g$ of promising configurations.
Using~\eqref{eq:p_value},
the estimated $p$-value associated
with configuration $\boldsymbol{\lambda}_j$ is computed by
\begin{equation}
\label{eq:p_val_def_max}
p^{\mathrm{opt}}_j
= \displaystyle\max_{\textrm{x}\in\{\mathrm{MC},\,\mathrm{MD}\}}
p\!\left(\{L^{\textrm{x}}_{i}(\boldsymbol{\lambda}_j)\}_{i\in\mathcal{D}_\mathrm{opt}};\,\alpha_{\textrm{x}}\right)
\end{equation}
Since neither loss function is bounded by $1$, the $p$-values are evaluated using the~\ac{WSR} bound~\cite{wsr} defined in~\eqref{eq:wsr_pvalue} (see Appendix~\ref{appendix:p_values}).

To identify the rejection set~\(\hat{\boldsymbol{\Lambda}}_{\mathrm{rej}}\!\subseteq\!\boldsymbol{\Lambda}_{\mathrm{sort}}\),~\ac{MHT} procedure is performed on the testing subset $\mathcal{D}_{\mathrm{tst}}$. For each configuration 
$\boldsymbol{\lambda}_j\!\in\!\boldsymbol{\Lambda}_{\mathrm{sort}}$, the null hypothesis
\begin{equation}
\label{eq:null_hypothesis_mc_md}
\mathcal{H}_j:
\bigl\{\mathcal{R}^{\mathrm{MD}}({\boldsymbol{\lambda}_j}) > \alpha_{\mathrm{MD}} \bigr\}
\;\lor\;
\bigl\{\mathcal{R}^{\mathrm{MC}}({\boldsymbol{\lambda}_j}) > \alpha_{\mathrm{MC}} \bigr\}
\end{equation}
is tested using a $p$-value $p^{\mathrm{tst}}_j$ defined analogously to~\eqref{eq:p_val_def_max}, but evaluated on $\mathcal{D}_{\mathrm{tst}}$.
The~\ac{FST} procedure is applied, stopping at the first index $j$ such that~$p^{\mathrm{tst}}_j\!>\!\delta$, yielding the rejection set~$\hat{\boldsymbol{\Lambda}}_{\mathrm{rej}}$.
Finally, a secondary filtering step is applied to~$\hat{\boldsymbol{\Lambda}}_{\mathrm{rej}}$ with respect to the free objective functions (\ac{FA} and~\ac{PA}) producing the final validated Pareto-optimal set~$\boldsymbol{\Lambda^*}\subseteq\hat{\boldsymbol{\Lambda}}_{\mathrm{rej}}$. Algorithm~\ref{algo:pareto_testing} summarizes the complete procedure, termed \textbf{`PT-SSL-U'} (Pareto-Testing for~\ac{SSL} with an unknown source count).
Fig.~\ref{fig:PT-SSL-U-flow} and Fig.~\ref{fig:inference-flow} illustrate the calibration and inference procedures, respectively.
As Pareto-Testing satisfies~\eqref{eq:alpha_delta_fwer} by design, every configuration~$\boldsymbol{\lambda}\!\in\!\boldsymbol{\Lambda}^*$ simultaneously controls the~\ac{MC} and~\ac{MD} risks with high probability and thus constitutes a plausible solution to~\eqref{eq:optimization_problem_final}. A practitioner can then choose any~$\boldsymbol{\lambda}\in\boldsymbol{\Lambda}^*$, for example by minimizing~\ac{PA},~\ac{FA}, or a weighted combination of the two.\looseness=-1

\textcolor{black}{
We conclude by emphasizing the main advantage of PT-SSL-U in the context of~\ac{SSL}. Assuming~$N$ discrete values per threshold, the~\ac{MOO} problem in~\eqref{eq:optimization_problem_final} induces a discretized configuration space of size~$|\boldsymbol{\Lambda}g|=N^{K{\mathrm{max}}+1}$, which may be prohibitively large for conventional sequential testing procedures and thereby reduce their statistical power. Moreover, testing the entire space may retain configurations that satisfy the controlled~\ac{MC} and~\ac{MD} risks but are suboptimal with respect to the free~\ac{FA} and~\ac{PA} objectives. To mitigate this, PT-SSL-U restricts hypothesis testing to the estimated Pareto-optimal set~$\boldsymbol{\Lambda}_{\mathrm{par}}\subseteq\boldsymbol{\Lambda}_g$, retaining configurations that represent the best trade-offs among the~\ac{MC},~\ac{MD},~\ac{FA}, and~\ac{PA} objectives. It then identifies the configurations for which the~\ac{MC} and~\ac{MD} constraints are statistically certified as feasible. Figure~\ref{fig:PT-SSL-U-flow} illustrates this procedure.
}
\begin{algorithm}[t]
\setstretch{1.3}  
\caption{PT-SSL-U: Pareto-Testing for~\ac{SSL}, unknown $K$}
\label{algo:pareto_testing}
\begin{algorithmic}[1]
\Require 
Calibration set $\mathcal{D}_{\mathrm{cal}}$; 
Discrete configuration set $\boldsymbol{\Lambda}_g$; Risk tolerance $\boldsymbol{\alpha}=(\alpha_{\mathrm{MD}}, \alpha_{\mathrm{MC}})$; Error significance $\delta$.
\Ensure Optimal risk-controlling set $\boldsymbol{\Lambda}^*$
\State $(\mathcal{D}_{\mathrm{opt}},\mathcal{D}_{\mathrm{tst}})
\gets \textsc{Split}(\mathcal{D}_{\mathrm{cal}})$
\Statex \hspace*{-\algorithmicindent}%
\textit{// Optimization:}  
\State $\{\widehat{\mathcal{R}}^\textrm{x}_n(\boldsymbol{\lambda})\}_{\textrm{x}\in\{\textrm{MD},\,\textrm{MC},\,\textrm{FA},\,\textrm{PA}\},\boldsymbol{\lambda}\in\boldsymbol{\Lambda_\textrm{g}}}
\gets \textsc{Risks}(\boldsymbol{\Lambda}_{\mathrm{g}}, \mathcal{D}_{\mathrm{opt}})$
\State $\boldsymbol{\Lambda}_{\mathrm{par}}
    \!\leftarrow\!\textsc{Pareto}
    \bigl(\{\widehat{\mathcal{R}}^\textrm{x}_n(\boldsymbol{\lambda})\}_{\textrm{x}\in\{\textrm{MD},\,\textrm{MC},\,\textrm{FA},\,\textrm{PA}\},\boldsymbol{\lambda}\in\boldsymbol{\Lambda_\textrm{g}}}\bigr)$     
\State $p^{\mathrm{opt}}_j \gets 
\displaystyle\max_{\textrm{x}\in\{\mathrm{MC},\,\mathrm{MD}\}}
p\!\left(\{L^{\textrm{x}}_{i}(\boldsymbol{\lambda}_j)\}_{i\in\mathcal{D}_\mathrm{opt}};\,\alpha_{\textrm{x}}\right), \forall\hfill\boldsymbol{\lambda}_j \in \boldsymbol{\Lambda}_{\mathrm{par}}$
\State $\boldsymbol{\Lambda}_{\mathrm{sort}}
   \gets \bigl(
        \boldsymbol{\lambda}_j:
        j=
        \operatorname*{\textrm{argsort}\>\>}\limits_{\boldsymbol{\lambda}_{q}\in\boldsymbol{\Lambda}_{\mathrm{par}}}p^\textrm{opt}_q
     \bigr)$\hfill \Comment{Ascending order}     
\Statex \hspace*{-\algorithmicindent}%
\textit{// Hypothesis Testing:}
\State $p^{\mathrm{tst}}_j \gets 
\displaystyle\max_{\textrm{x}\in\{\mathrm{MC},\,\mathrm{MD}\}}
p\!\left(\{L^{\textrm{x}}_{i}(\boldsymbol{\lambda}_j)\}_{i\in\mathcal{D}_\mathrm{tst}};\,\alpha_{\textrm{x}}\right),\> \forall\boldsymbol{\lambda}_j \!\in\! \boldsymbol{\Lambda}_{\mathrm{sort}}$
\State $J \gets
\min\bigl\{ 1\leq j\leq|\boldsymbol{\Lambda}_\textrm{sort}| :
      p^{\mathrm{tst}}_j > \delta
\bigr\}$
\hfill\Comment{FST procedure}

\State $\hat{\boldsymbol{\Lambda}}_{\mathrm{rej}}
    \gets \{\boldsymbol{\lambda}_j: j < J\}$
\State $\{\widehat{\mathcal{R}}^\textrm{x}_n(\boldsymbol{\lambda})\}_{\textrm{x}\in\{\textrm{MD},\,\textrm{MC},\,\textrm{FA},\,\textrm{PA}\},\boldsymbol{\lambda}\in\boldsymbol{\hat{\Lambda}_\textrm{rej}}} \gets \textsc{Risks}(\hat{\boldsymbol{\Lambda}}_{\mathrm{rej}}, \mathcal{D}_{\mathrm{tst}})$
\State $\boldsymbol{\Lambda}^{*}
    \!\gets\! \textsc{Pareto}
    \bigl(\{\widehat{\mathcal{R}}^\textrm{x}_n(\boldsymbol{\lambda})\}_{\textrm{x}\in\{\textrm{FA},\,\textrm{PA}\},\boldsymbol{\lambda}\in\boldsymbol{\hat{\Lambda}_\textrm{rej}}}\bigr)$
\State \Return $\boldsymbol{\Lambda}^{*}$
\end{algorithmic}
\end{algorithm}

\subsection{Computational Complexity}
\label{subsec:complexity}
\BL{We discuss the computational complexity of our proposed frameworks. It is important to emphasize that they operate as post-hoc layers on top of a likelihood-map-based~\ac{SSL} backend. Accordingly, likelihood map construction and iterative source localization, as described in Algorithm~\ref{alg:iterative_ssl}, are performed before applying the proposed risk-control mechanisms.
The additional computational burden is mainly concentrated in the offline calibration stage, where the risk-control thresholds are selected using calibration data. Once these thresholds are fixed, online inference during test time only requires source detection followed by prediction-region construction, as described in Algorithm~\ref{algo:prediction_set}.
Notably, even in highly reverberant or ambiguous scenes, where the prediction region may form a large connected component, the cost of region construction remains bounded by $\mathcal{O}(|\mathcal{G}|)$, since each grid point is visited at most once. The resulting computational complexities are summarized in Table~\ref{tab:complexity}. For simplicity, fixed multiplicative constants such as $K$, $K_{\max}$, and the number of objectives are omitted from the asymptotic notation. Procedures whose complexity is dominated by the leading-order terms are also omitted.}\looseness=-1

\begin{table*}[t]
\color{black}
\centering
\caption{Computational complexity of the main components.}
\label{tab:complexity}
\scriptsize
\setlength{\tabcolsep}{5pt}
\renewcommand{\arraystretch}{1.7}
\begin{tabular}{@{}p{0.25\textwidth}p{0.26\textwidth}@{\hspace{10pt}}c@{\hspace{8pt}}c@{\hspace{12pt}}p{0.28\textwidth}@{}}
\toprule
Component 
& Time complexity 
& Offline
& Online
& Comment \\
\midrule

Likelihood-map construction/update.~\eqref{eq:likelihood_map},~\eqref{eq:srp-dnn-likelihood-map}
& \makecell[l]{
$C_{\Phi}\!=\!\mathcal O(|\mathcal G|QFT)$ for~\ac{SRP-PHAT} \\
$C_{\Phi}\!=\!\mathcal O(C_{\mathrm{NN}}+|\mathcal G|QFT)$ for SRP-DNN
}
& \cmark
& \cmark
& \makecell[l]{
$\mathcal{O}(C_{\mathrm{NN}})$ for feedforward cost of SRP-DNN.
} \\

Iterative source detection and localization.~Algorithm~\ref{alg:iterative_ssl}
& $\mathcal O\!\left(C_{\Phi}+|\mathcal G|\right)$
& \cmark
& \cmark
& Second term is for peak search.\\

Prediction-region construction.~Algorithm~\ref{algo:prediction_set}
& $\mathcal O(|\mathcal G|)$
& \cmark
& \cmark
& \makecell[l]{
Per sample, source, and threshold value.
} \\

CRC-SSL-N calibration
& \makecell[l]{
$\begin{aligned}
&\mathcal O\!\left(n\!\left(C_{\Phi}+|\Lambda^{\mathrm{MC}}||\mathcal G|\right)\!+\!|\Lambda^{\mathrm{MC}}|\!\log |\Lambda^{\mathrm{MC}}|\right)
\end{aligned}$
}
& \cmark
& \xmark
& \makecell[l]{
The first term constructs calibration regions; \\
the second
term computes the empirical\\$1-\alpha$-quantile to select $\hat{\lambda}^{\mathrm{MC}}_k$.
}
\\

CRC-SSL-N test
& $\mathcal O\!\left(C_{\Phi}+|\mathcal G|\right)$
& \xmark
& \cmark
& \makecell[l]{Per test sample
} \\

PT-SSL-U calibration.~Algorithm~\ref{algo:pareto_testing}
& \makecell[l]{$\mathcal O\!\left(n(C_{\Phi}+|\Lambda^{\mathrm{MC}}||\mathcal G|)+n|\boldsymbol{\Lambda}_g|+|\boldsymbol{\Lambda}_g|^2\right)$}
& \cmark
& \xmark
& \makecell[l]{
$n|\boldsymbol{\Lambda}_g|$ evaluates empirical risks over the config-\\uration grid; $|\boldsymbol{\Lambda}_g|^2$ recovers the Pareto frontier.
} \\

PT-SSL-U test
& $\mathcal O\!\left(C_{\Phi}+|\mathcal G|\right)$
& \xmark
& \cmark
& \makecell[l]{Per test sample
} \\

\bottomrule
\end{tabular}
\vspace{-0.4cm}
\end{table*}

\section{Experimental Study}
\label{sec:simulation_and_results}
We evaluate the proposed approaches on simulated data as well as real-world recordings, using the~\ac{SRP-PHAT} algorithm and the SRP-DNN model to generate 2D spatial likelihood maps. Experiments cover diverse reverberation times, static and moving sources, and varying numbers of sources, under both known- and unknown-source-count settings.\looseness=-1

\subsection{Datasets}
\label{subsec:epx_setup}
Throughout the experiments, we employ a~$12$-element pseudo-spherical microphone array, consistent with the array configuration used in the LOCATA challenge~\cite{lollmann2018locata}. Each detected source is characterized by its two-dimensional~\ac{DOA}~$\mathbf{p}_i \!= \![\phi_i,\, \theta_i]^{\mathsf{T}}$, where~$\phi_i \!=\! 0$ and $\theta_i \!=\! 0$
align with the positive $z$- and $x$-axes, respectively.
\textcolor{black}{The 2D~\ac{DOA} grid is discretized with an angular resolution of~\SI{5}{\degree} in both dimensions. 
In turn, the smallest non-empty region corresponds to approximately~$0.04\%$ of the grid. Accordingly, very small PA values should be interpreted with this discretization limit in mind. Finer grids would reduce the minimum non-zero~\ac{PA} and result in more compact regions.}
For all experiments, the attenuation factor in Algorithm~\ref{alg:iterative_ssl} is set to $\gamma=0.5$.
\looseness=-1

The SRP-DNN model is trained following~\cite{srp_dnn}, using randomized room dimensions, microphone-array positions, reverberation times, and~\ac{SNR} levels. 
Below, we describe the data used for calibration and evaluation. All signals are analyzed in the~\ac{STFT} domain using a Hann window with~$512$ frequency bins and~$50$\% frame overlap.
For both the~\ac{SRP-PHAT} and SRP-DNN simulations, we use the implementation provided in~\cite{srp_dnn}. For static sources, the spatial likelihood maps defined in~\eqref{eq:likelihood_map} and~\eqref{eq:srp-dnn-likelihood-map} are averaged over all time frames, whereas for moving sources, they are computed separately for each time frame.
\looseness=-1

\subsubsection{\textbf{Synthetic}}
This dataset is used for both calibration and testing. Clean speech utterances are randomly drawn from the LibriSpeech corpus~\cite{panayotov2015librispeech}. \acp{RIR} are generated based on the image-source method implemented in~\cite{diaz2021gpurir}. Various simulated rooms were considered, with reverberation time set to~$T_{60} \!\in\! \{400, 700,1000\}\,\mathrm{ms}$ and fixed dimensions of~\SI{6}{m}~$\times$~\SI{6}{m}~$\times$~\SI{2.5}{m}.
The microphone signals are obtained by convolving each speech signal with the corresponding~\ac{RIR}, while adding~\ac{AWGN} with~\ac{SNR} of~$\{5,10,15\}\,\si{dB}$. The maximum number of sources is set to~$K_{\max}\!=\!3$. 

\textcolor{black}{Throughout our experiments, we assume a reasonable minimum inter-source angular separation of~$\SI{15}{\degree}$ in both azimuth and elevation. Accordingly, the detector uses $d\!=\!\SI{15}{\degree}$ as the minimum separation parameter between peaks, which corresponds to three grid steps with our map resolution of~$\SI{5}{\degree}$. This detector-level parameter does not, however, restrict the applicability of the proposed frameworks, which can accommodate any prescribed separation distance~$d$. Overall, this parameter should be set according to the application and the assumed minimal separation between sources.}
\looseness=-1

For the \textcolor{black}{known-source-count setting}, we use~\(500\) calibration samples and~\(100\) test samples per source-count~\(K \!\in\! \{1,\ldots,K_{\max}\}\). For the \textcolor{black}{unknown-source-count setting}, we generate~\(1500\) samples (\(500\) for each source count~\(K \!\in\! \{1,2,3\}\)) and partition them into~\(1200\) calibration samples and \(300\) test samples, allocated equally across the three~\(K\) values.

\subsubsection{\textbf{LOCATA}}
We use real-world speech recordings from the LOCATA challenge dataset~\cite{lollmann2018locata}.
The recordings were collected in a room of size $\SI{7.1}{m}\times\SI{9.8}{m}\times\SI{3}{m}$ and a measured reverberation time of $T_{60}=\SI{550}{ms}$.

We consider two LOCATA evaluation settings:
(i) static sources, using a single test set comprising~\texttt{Task-1}, which contains recordings of a single active source, and~\texttt{Task-2}, which includes scenarios with two static sources; \textcolor{black}{and
(ii) moving sources, using~\texttt{Task-3} and~\texttt{Task-4}, corresponding to single- and two-moving-source recordings, respectively.}
For the static-source evaluation, due to the limited number of two-source static scenarios in~\texttt{Task-2}, additional samples are generated by mixing
single-source recordings from~\texttt{Task-1}. The test set consists of~$20$ LOCATA recordings, comprising~$10$ single-source and~$10$ two-source scenarios. Calibration is carried out using LOCATA-matched synthetic data (described below).
\textcolor{black}{For the moving-source evaluation, we consider both known- and unknown-source-count test cases. As in~\cite{srp_dnn}, each sample corresponds to a single time frame. We use a total of~$|\mathcal{D}_{\mathrm{cal}}|=1400$ samples, equally split between~$K\!\in\!\{1,2\}$, with~$|\mathcal{D}_{\mathrm{opt}}|=|\mathcal{D}_{\mathrm{tst}}|=650$ optimization and hypothesis-testing samples, respectively, and~$100$ test samples.}
\looseness=-1

\subsubsection{\textbf{LOCATA-Matched Synthetic}}
This dataset serves two purposes. It is used for calibration prior to evaluation on real-world LOCATA recordings, and it acts as a synthetic baseline for comparative performance assessment.
It mirrors the LOCATA recording environment with $K_{\max}\!=\!2$. We use $400$ samples for calibration and optimization in the Pareto-Testing procedure.

\subsection{Calibration Process}
Given a set of true~\acp{DOA}
$\{(\mathbf{p}^*_{i,1}, \ldots, \mathbf{p}^*_{i,K})\}_{i=1}^n$,
a corresponding acoustic scene is simulated for each realization.
The~\ac{SRP-PHAT} or SRP-DNN models are then applied to generate an initial 2D spatial likelihood map~$\Phi_i$. We describe the calibration process in each setting. 
\subsubsection{Known-source count}
With $K$ known, the calibration process targets the coverage metric and is performed separately for each source $k \!\in\! \{1,\ldots,K\}$.
Algorithm~\ref{alg:iterative_ssl} is run for exactly $K$ iterations, yielding the estimated directions together with their associated likelihood maps.  
The calibration set for the $k$th detected source in the $K$-source scenario is defined as
\begin{equation}
    \label{eq:d_cal_K_k}
    \mathcal{D}^{(k)}_{\mathrm{ cal},K}=\left\{ \left((\Phi_i^{(k)}, \hat{\mathbf{p}}_{i,k}), \mathbf{p}^*_{i,k}\right) \right\}_{i\in\mathcal{D}_{\mathrm{cal},K}}.
\end{equation}
For each source $k$, the threshold space $\Lambda^{\mathrm{MC}}_k\! =\! [0,1]$ is discretized into $100$ equally spaced values.
Prediction regions $\mathcal{C}_{\lambda_k}(\hat{\mathbf{p}}_{i,k})$ are constructed using Algorithm~\ref{algo:prediction_set}, and the optimal threshold $\hat{\lambda}_k$ is selected according to~\eqref{eq:crc_lambda_hat}.

\subsubsection{Unknown-source count}
Here, we perform joint detection and localization, with the objective of selecting configurations~$\hat{\boldsymbol{\lambda}} = (\hat{\lambda}_1, \ldots, \hat{\lambda}_{K_\mathrm{max}}, \hat{\beta})$ that solve the constrained~\ac{MOO} defined in~\eqref{eq:optimization_problem_final}. Assume that $K_i$ denotes the number of active sources in the $i$th recording. The corresponding calibration sample consists of $K_{\mathrm{max}}$ pairs of predicted source locations and their associated spatial likelihood maps,
together with the ground-truth source locations
\begin{equation}
\label{eq:D_k_cal}
\mathcal{D}_{\mathrm{cal},K_{\max}}
\!=\!\left\{\!\left(
\{(\Phi_i^{(k)}, \hat{\mathbf{p}}_{i,k})\}_{k=1}^{K_{\max}},
\{\mathbf{p}_{i,k}^*\}_{k=1}^{K_i}
\right) \!\right\}_{i\in\mathcal{D_{\mathrm{cal}}}}.
\end{equation}
\textcolor{black}{Similarly to the known-source-count setting, prediction regions are constructed iteratively for each detected peak
using Algorithm~\ref{algo:prediction_set}.\looseness=-1}

Since the scale of $\Phi$ varies across recordings due to noise realizations and speech variability, each map is linearly normalized to the interval $[0,1]$. We define a discrete configuration grid
$\boldsymbol{\Lambda}_g \subset [0,1]^{K_{\mathrm{max}}+1}$, 
using $N\!=\!15$ values per dimension for $K_{\max}\!=\!3$ and $N\!=\!35$ for $K_{\max}\!=\!2$,
resulting in $N^{K_{\max}+1}$ candidate configurations.\looseness=-1

Finally, we note that performing calibration on the LOCATA-matched synthetic dataset and evaluating on LOCATA real recordings requires care, as the two are not exchangeable, violating a key assumption of~\ac{CRC} and~\ac{LTT}, as discussed in Appendix~\ref{appndx:covariate_shift}. This induces a distribution shift in detected peak values, as shown in Fig.~\ref{fig:cov_shift}. While methods for handling non-exchangeable settings in~\ac{CP} have been proposed~\cite{cp_cov_shift}, they are beyond the scope of this work. Instead, we adopt a simpler mitigation strategy that aligns the peak likelihood distributions (see Appendix~\ref{appndx:covariate_shift}).\looseness=-1

\subsection{Experiments and Performance Evaluation}
\label{subsec:exp_and_performance_eval}
We present the results for each scenario, reporting~\ac{MC}, \ac{PA}, and azimuth/elevation-averaged~\ac{MAE}. The metrics are averaged over~$100$ random partitions into optimization and hypothesis-testing sets for PT-SSL-U, or a calibration set for CRC-SSL-N, together with a test set.\looseness=-1

\subsubsection{\textbf{Known-source count}}
In this setting, the objective is to control the~\ac{MC} risk at a prescribed level~$\alpha_{\mathrm{MC}}$, as defined in~\eqref{eq:mc_risk_control_known_speakers_amount}, separately for each source~$k \!\in\!\{1,\ldots,K\}$.
We consider significance levels~$\alpha_{\mathrm{MC}} \!\in\! \{0.1, 0.05\}$, corresponding to target coverage levels of $90\%$ and $95\%$, respectively.

Table~\ref{tab:CRC_known_source_merged} summarizes the results obtained by applying~\textcolor{black}{CRC-SSL-N} to spatial likelihood maps generated by~\ac{SRP-PHAT} and SRP-DNN. For both methods, the empirical coverage closely matches the prescribed nominal levels across all source counts and room conditions.
A consistent trend is observed whereby higher target coverage (from $90\%$ to $95\%$) and larger $T_{60}$ values lead to systematically larger regions, reflecting increased uncertainty under stricter reliability requirements and more challenging acoustic conditions.
This effect is most pronounced for~\ac{SRP-PHAT}. While the first detected source is localized with high confidence and compact regions, later detected sources show expanding prediction regions as reverberation increases, indicating a substantial loss of localization certainty.
In contrast, SRP-DNN exhibits more stable behavior: prediction regions remain largely compact, with only mild growth as coverage level and reverberation increase. 

\BL{
We further evaluate CRC-SSL-N under more challenging acoustic conditions, with~$T_{60}\!=\!\SI{1000}{ms}$ and~\ac{SNR} values of $\{5,10\}\,\si{dB}$. We focus on SRP-DNN only, since~\ac{SRP-PHAT} already produces substantially larger prediction regions under milder acoustic conditions.
The prescribed~\ac{MC} levels are still tightly achieved. However, the~\ac{PA} increases noticeably, especially for the last detected source in the~$K\!=\!3$ multi-source setting.
The larger~\acp{PA}, accompanied by higher~\ac{MAE}, further reinforce \textcolor{black}{the role of the prediction-region area as a meaningful~\ac{UQ} indicator}; see also Section~\ref{sec:comp_with_baselines} for further discussion.
}\looseness=-1

\begin{figure*}[!]
    \centering

    \begin{minipage}{0.3\textwidth}
        \centering
        \includegraphics[width=\linewidth]{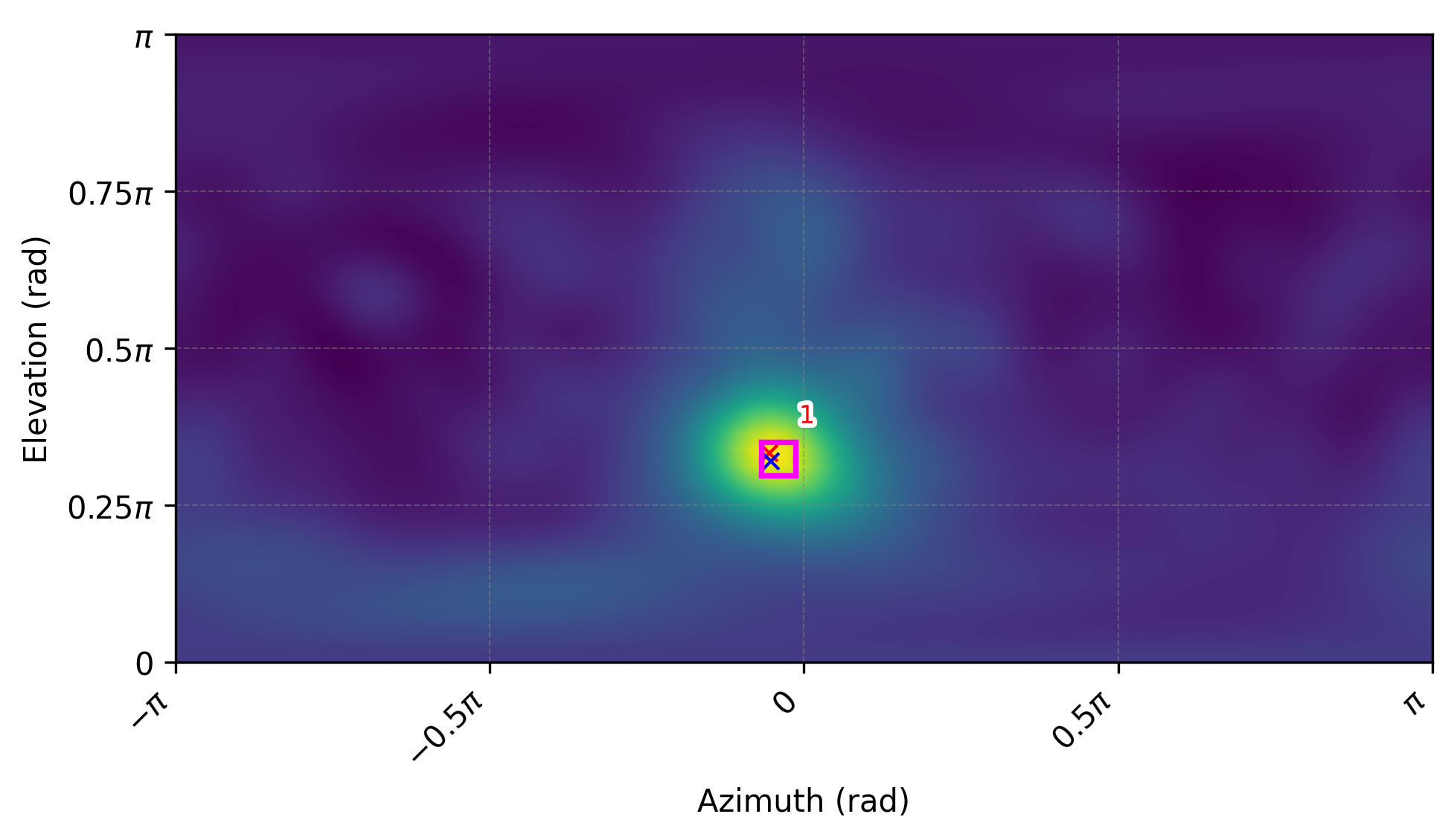}
    \end{minipage}
    \hfill
    \begin{minipage}{0.3\textwidth}
        \centering
        \includegraphics[width=\linewidth]{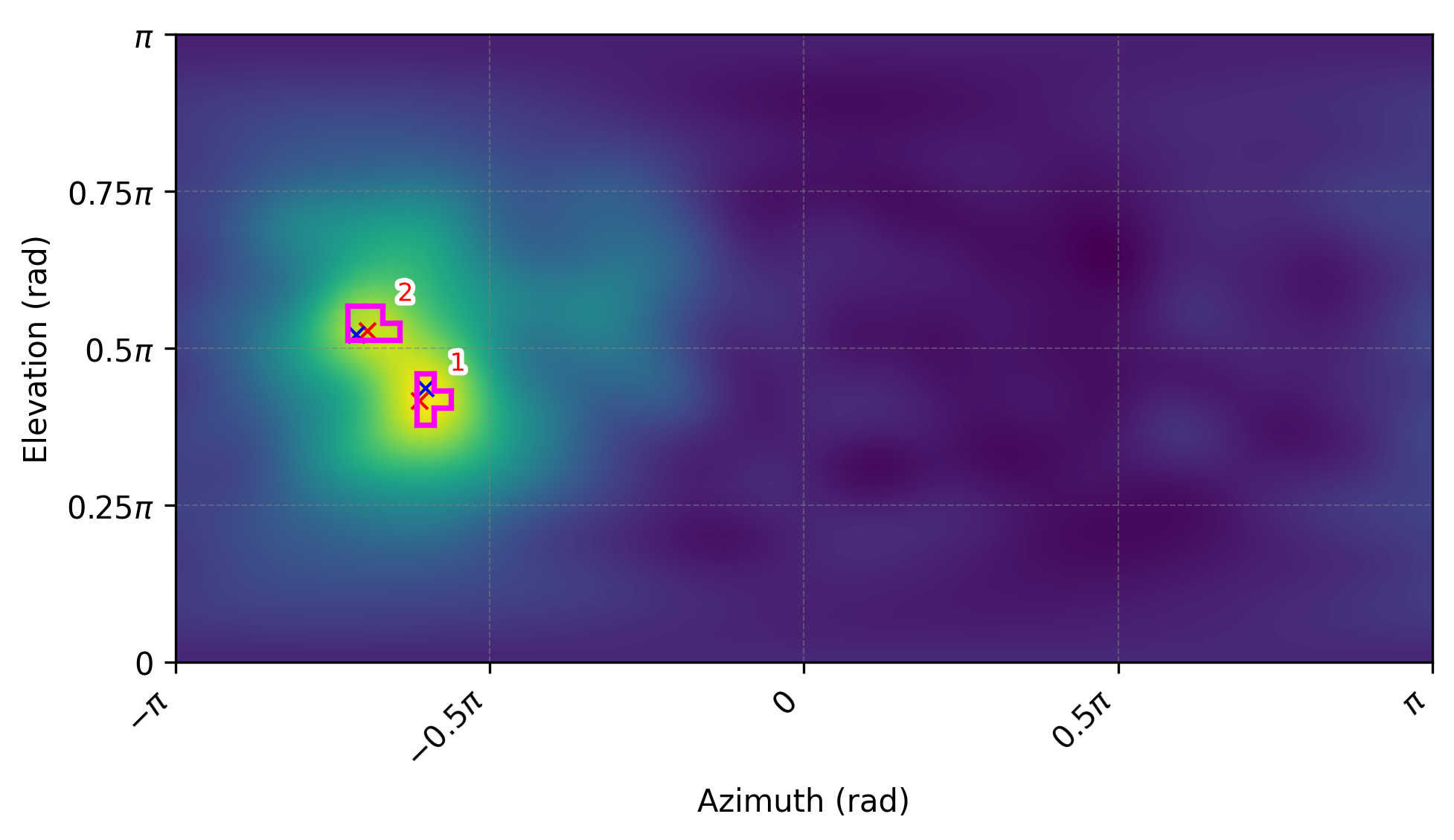}
    \end{minipage}
    \hfill
    \begin{minipage}{0.3\textwidth}
        \centering
        \includegraphics[width=\linewidth]{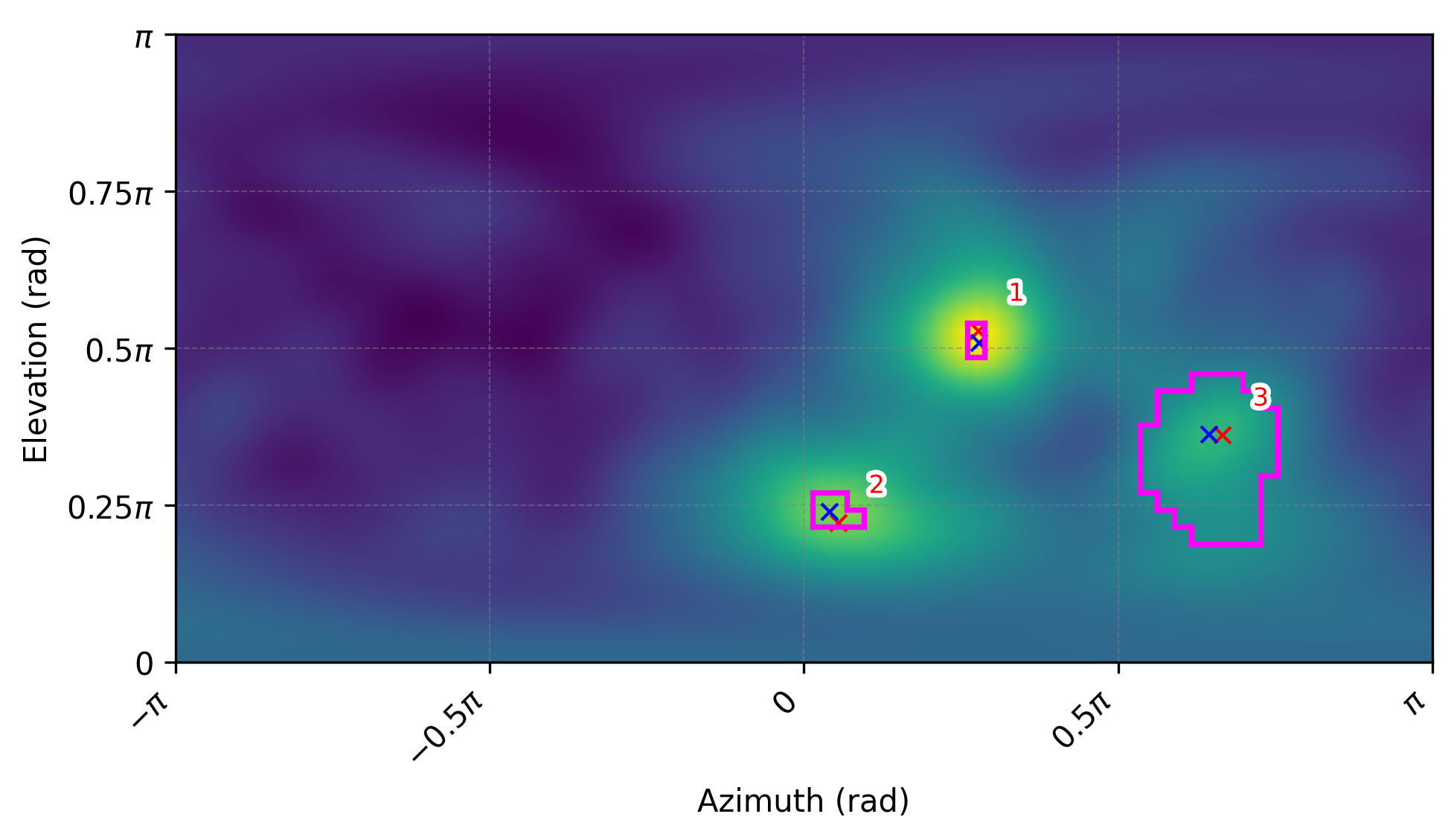}
    \end{minipage}

    \begin{minipage}{0.3\textwidth}
        \centering
        \includegraphics[width=\linewidth]{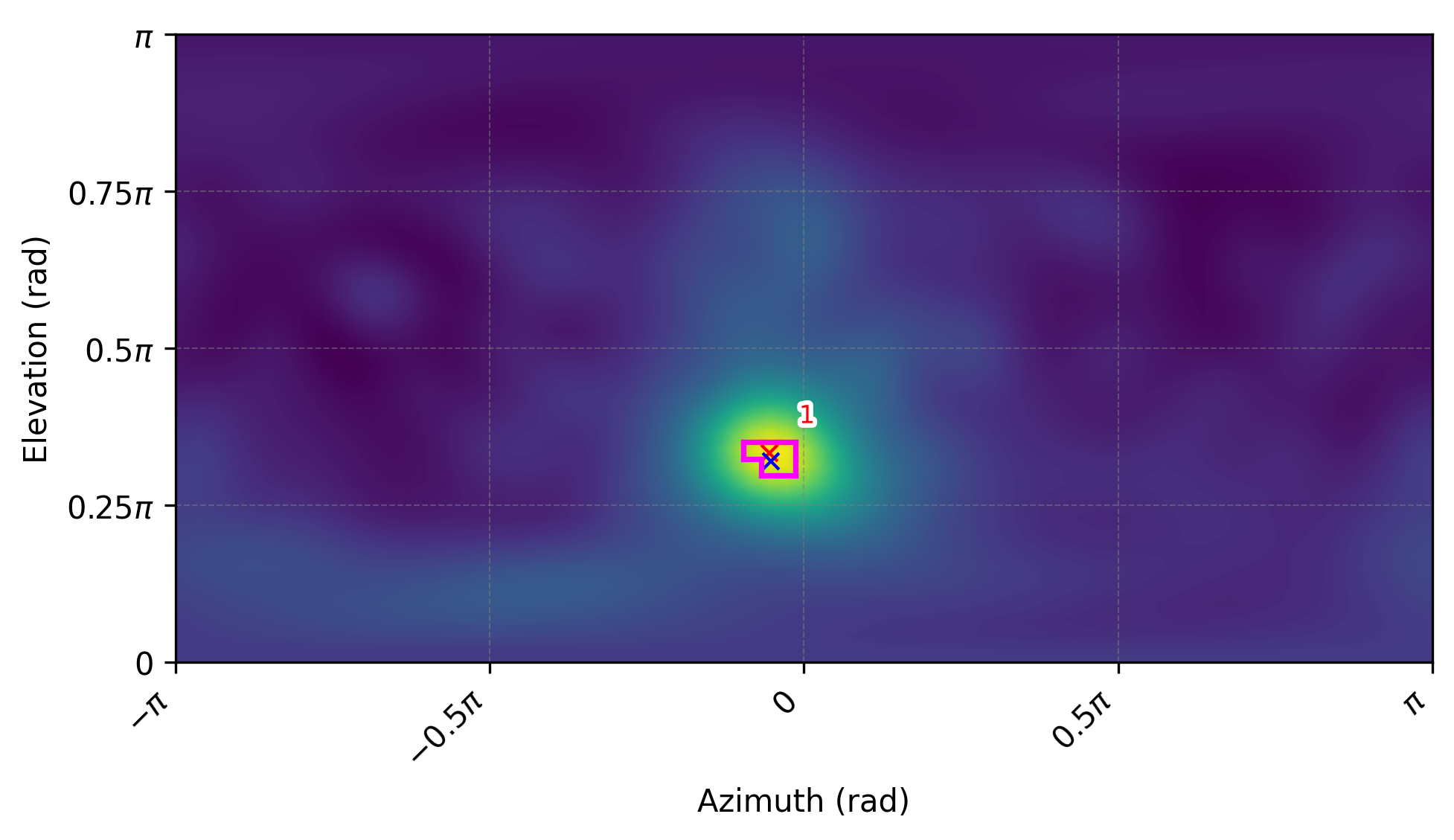}
    \end{minipage}
    \hfill
    \begin{minipage}{0.3\textwidth}
        \centering
        \includegraphics[width=\linewidth]{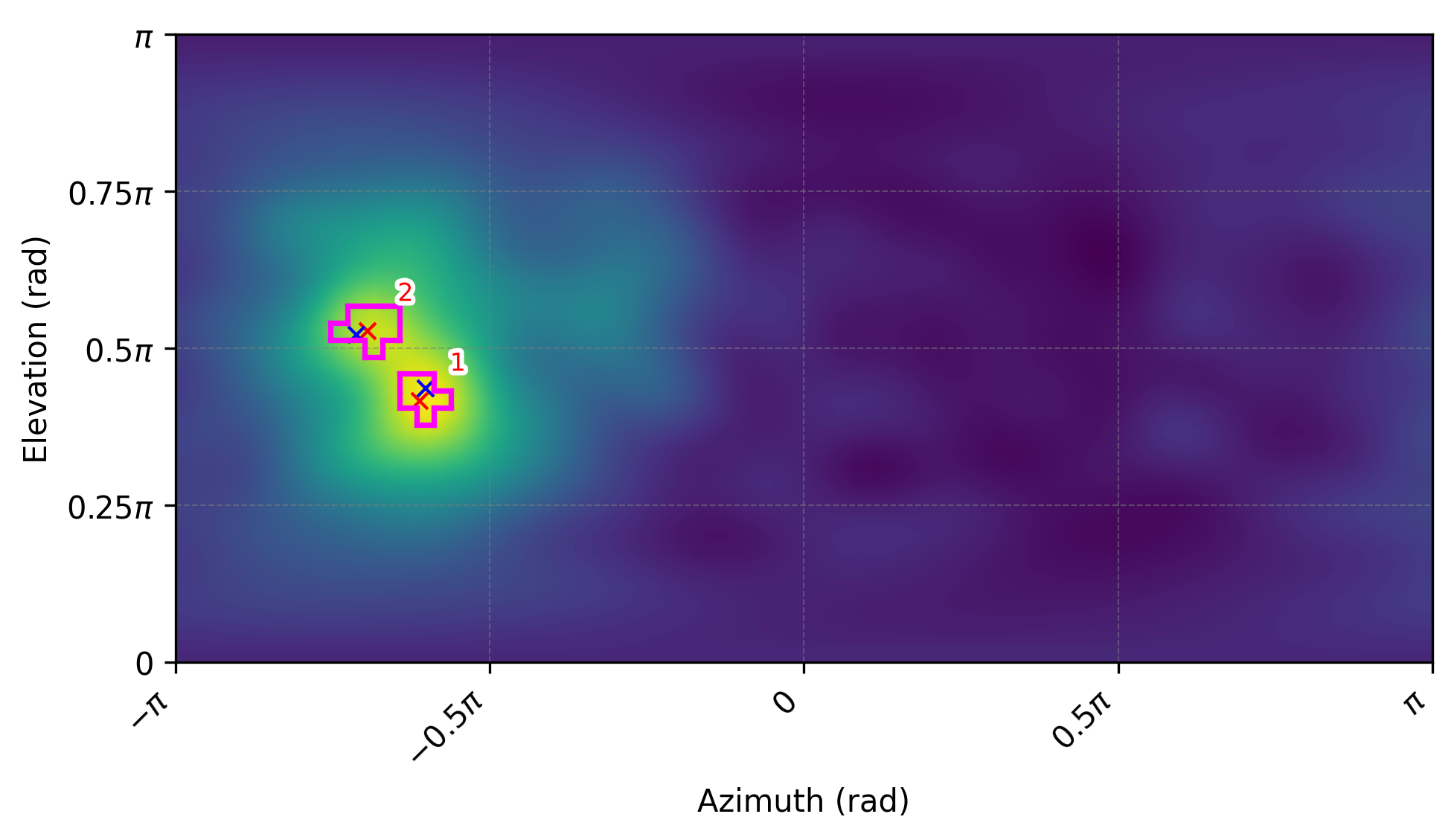}
    \end{minipage}
    \hfill
    \begin{minipage}{0.3\textwidth}
        \centering
        \includegraphics[width=\linewidth]{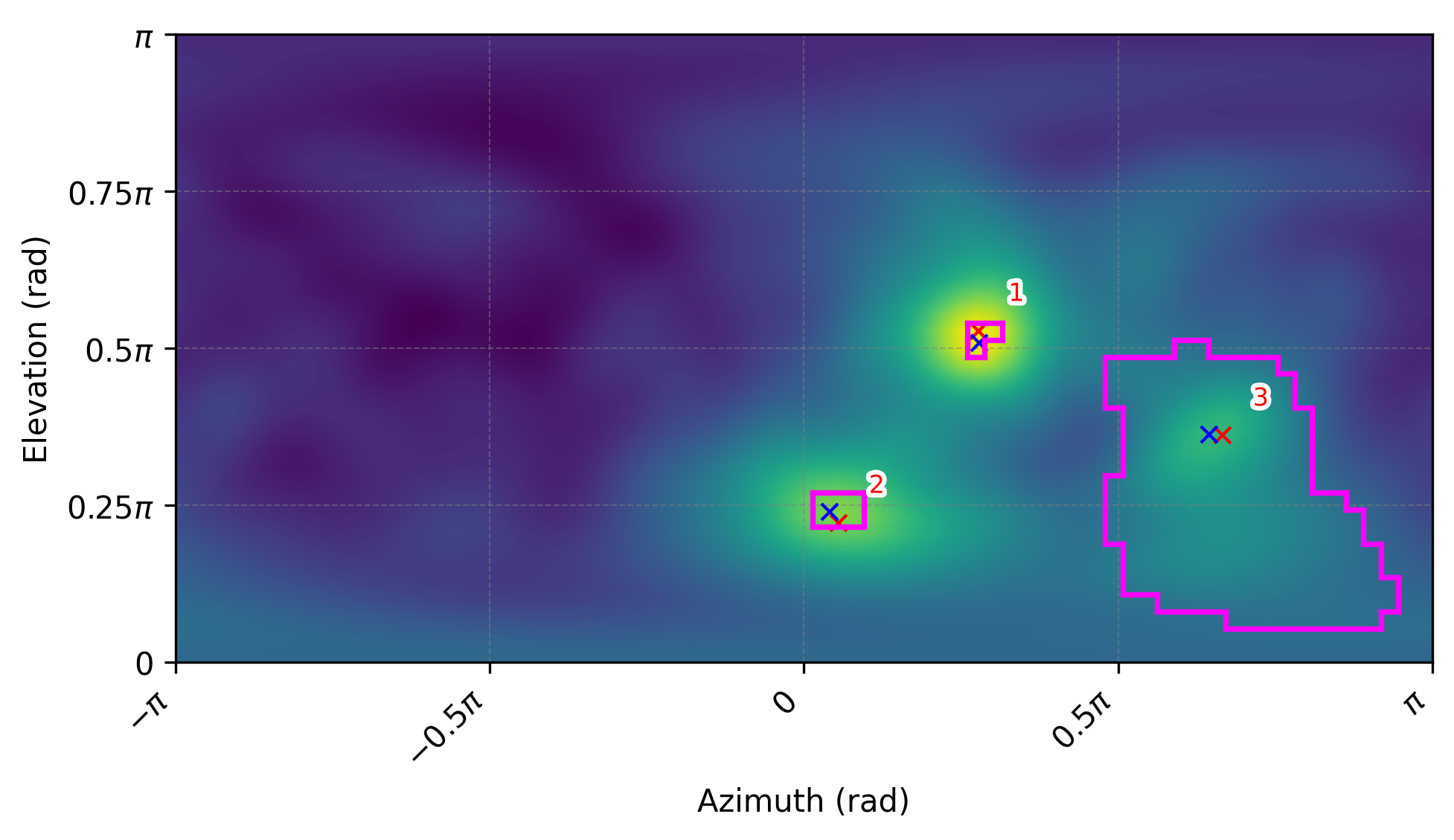}
    \end{minipage}

    \caption{
        CRC-SSL-N prediction regions applied to SRP-DNN likelihood maps at \SI{90}{\%} (top) and \SI{95}{\%} (bottom) coverage, for $T_{60}=700$\,ms, and \ac{SNR}$=15$\,dB. Blue and red Xs mark true and detected sources, respectively, ordered by detection. Brighter colors indicate higher likelihood. Regions expand adaptively along likelihood contours rather than fixed geometric shapes.\looseness=-1
    }
    \label{fig:coverage_columns_dnn}
    \vspace{-2em}
\end{figure*}

Fig.~\ref{fig:coverage_columns_dnn} shows the coverage regions obtained using the~\textcolor{black}{CRC-SSL-N} method applied to SRP-DNN likelihood maps for~$K\!\in\!\{1,2,3\}$ active sources under~$T_{60}\!=\!\SI{700}{ms}$ and $\mathrm{SNR}=15\,\mathrm{dB}$. The dominant source in each case has the strongest energy and an estimated~\ac{DOA} very close to the ground truth, yielding a highly compact prediction region. In contrast, the third source in the~$K\!=\!3$ scenario receives a noticeably larger region, despite appearing well localized in this sample, reflecting its weaker energy and the influence of residual errors from earlier source removals.\looseness=-1

Additional experimental results for the~\textcolor{black}{CRC-SSL-N} procedure are provided in Appendix~\ref{sec:app_b}. In Appendix~\ref{appndx:multi_room_crc}, it is applied jointly across all room settings (i.e., different reverberation levels), rather than on a per-room basis. This strategy is particularly relevant when only limited calibration data are available per room, or when the target room conditions are unknown and cannot be matched to a specific calibration set.
In Appendix~\ref{appndx:ltt_known_source_count},~\textcolor{black}{CRC-SSL-N} is replaced by the Pareto-Testing procedure in the known-source-count setting, providing stronger performance guarantees with high probability.

\begin{table*}[t]
\centering
\scriptsize
\caption{CRC-SSL-N results for static multi-\ac{SSL} with known $K\in\{1,2,3\}$, using \ac{SRP-PHAT} and SRP-DNN.}
\label{tab:CRC_known_source_merged}
\setlength{\tabcolsep}{0.8pt}
\renewcommand{\arraystretch}{1.}
\sisetup{table-number-alignment = center}
\newcommand{\NA}{\multicolumn{1}{c}{--}}

\begin{adjustbox}{max width=\textwidth}
\begin{tabular}{@{} c c c c *{18}{S[table-format=2.4]} @{}}
\toprule
\multirow{4}{*}{\makecell{}}
  & \multirow{4}{*}{\makecell{$T_{60}$\\$\mathrm{[ms]}$}}
  & \multirow{4}{*}{\makecell{SNR\\$\mathrm{[dB]}$}}
  & \multirow{4}{*}{$\alpha_{\mathrm{MC}}$}
  & \multicolumn{3}{c}{$K=1$}
  & \multicolumn{6}{c}{$K=2$}
  & \multicolumn{9}{c}{$K=3$} \\
\cmidrule(lr){5-7} \cmidrule(lr){8-13} \cmidrule(lr){14-22}
  &  &  &  
  & \multicolumn{3}{c}{Source 1}
  & \multicolumn{3}{c}{Source 1} & \multicolumn{3}{c}{Source 2}
  & \multicolumn{3}{c}{Source 1} & \multicolumn{3}{c}{Source 2} & \multicolumn{3}{c}{Source 3} \\
\cmidrule(lr){5-7} \cmidrule(lr){8-10} \cmidrule(lr){11-13}
\cmidrule(lr){14-16} \cmidrule(lr){17-19} \cmidrule(lr){20-22}
  &  &  &  
  & \multicolumn{1}{c}{\ac{MC}} & \multicolumn{1}{c}{PA\si{\%}} & \multicolumn{1}{c}{\makecell{MAE\si{\degree}}}
  & \multicolumn{1}{c}{\ac{MC}} & \multicolumn{1}{c}{PA\si{\%}} & \multicolumn{1}{c}{\makecell{MAE\si{\degree}}}
  & \multicolumn{1}{c}{\ac{MC}} & \multicolumn{1}{c}{PA\si{\%}} & \multicolumn{1}{c}{\makecell{MAE\si{\degree}}}
  & \multicolumn{1}{c}{\ac{MC}} & \multicolumn{1}{c}{PA\si{\%}} & \multicolumn{1}{c}{\makecell{MAE\si{\degree}}}
  & \multicolumn{1}{c}{\ac{MC}} & \multicolumn{1}{c}{PA\si{\%}} & \multicolumn{1}{c}{\makecell{MAE\si{\degree}}}
  & \multicolumn{1}{c}{\ac{MC}} & \multicolumn{1}{c}{PA\si{\%}} & \multicolumn{1}{c}{\makecell{MAE\si{\degree}}} \\
\midrule

\multirow{5}{*}{\makecell{SRP- \\ PHAT}}
  & \multirow{2}{*}{400}
    & \multirow{2}{*}{15}
    & 0.100 
    & 0.102 & 0.15 & \multicolumn{1}{c}{\multirow{2}{*}{$2.4$}}  
    & 0.098 & 0.16 & \multicolumn{1}{c}{\multirow{2}{*}{$2.2$}} 
    & 0.105 & 18.89 & \multicolumn{1}{c}{\multirow{2}{*}{$15.8$}} 
    & 0.097 & 0.16 & \multicolumn{1}{c}{\multirow{2}{*}{$2.2$}} 
    & 0.100 & 4.25 & \multicolumn{1}{c}{\multirow{2}{*}{$9.8$}} 
    & 0.098 & 57.63 & \multicolumn{1}{c}{\multirow{2}{*}{$33.5$}} \\
  & & & 0.050 
    & 0.048 & 0.21 & 
    & 0.048 & 0.21 & 
    & 0.052 & 45.79 &  
    & 0.047 & 0.23 &  
    & 0.048 & 25.09 &  
    & 0.048 & 79.30 &  \\
\cmidrule(lr){2-22}
  & \multirow{2}{*}{700}
    & \multirow{2}{*}{15}
    & 0.100 
    & 0.094 & 0.19 & \multicolumn{1}{c}{\multirow{2}{*}{$2.7$}} 
    & 0.097 & 0.20 & \multicolumn{1}{c}{\multirow{2}{*}{$2.6$}} 
    & 0.106 & 31.43 & \multicolumn{1}{c}{\multirow{2}{*}{$19.0$}} 
    & 0.094 & 0.22 & \multicolumn{1}{c}{\multirow{2}{*}{$2.5$}} 
    & 0.096 & 12.66 & \multicolumn{1}{c}{\multirow{2}{*}{$12.6$}} 
    & 0.101 & 67.17 & \multicolumn{1}{c}{\multirow{2}{*}{$39.9$}} \\
  & & & 0.050 
    & 0.048 & 0.30 &  
    & 0.049 & 0.29 & 
    & 0.052 & 61.57 & 
    & 0.047 & 0.28 & 
    & 0.048 & 45.69 &  
    & 0.048 & 85.58 &  \\

\midrule

\multirow{10}{*}{\makecell{SRP- \\ DNN}}
  & \multirow{2}{*}{400}
    & \multirow{2}{*}{15}
    & 0.100 
    & 0.097 & 0.09 & \multicolumn{1}{c}{\multirow{2}{*}{$1.6$}} 
    & 0.092 & 0.09 & \multicolumn{1}{c}{\multirow{2}{*}{$1.6$}}
    & 0.101 & 0.15 & \multicolumn{1}{c}{\multirow{2}{*}{$2.7$}} 
    & 0.096 & 0.09 & \multicolumn{1}{c}{\multirow{2}{*}{$1.6$}}
    & 0.098 & 0.12 & \multicolumn{1}{c}{\multirow{2}{*}{$1.9$}} 
    & 0.101 & 1.62 & \multicolumn{1}{c}{\multirow{2}{*}{$9.2$}} \\
  & & & 0.050 
    & 0.052 & 0.12 &  
    & 0.044 & 0.12 & 
    & 0.051 & 0.19 & 
    & 0.045 & 0.13 & 
    & 0.047 & 0.17 &  
    & 0.049 & 12.17 &  \\
\cmidrule(lr){2-22}
  & \multirow{2}{*}{700}
    & \multirow{2}{*}{15}
    & 0.100 
    & 0.097 & 0.12 & \multicolumn{1}{c}{\multirow{2}{*}{$1.8$}}
    & 0.093 & 0.12 & \multicolumn{1}{c}{\multirow{2}{*}{$1.7$}}
    & 0.098 & 0.19 & \multicolumn{1}{c}{\multirow{2}{*}{$3.0$}}
    & 0.098 & 0.11 & \multicolumn{1}{c}{\multirow{2}{*}{$1.8$}} 
    & 0.093 & 0.17 & \multicolumn{1}{c}{\multirow{2}{*}{$2.2$}} 
    & 0.098 & 3.87 & \multicolumn{1}{c}{\multirow{2}{*}{$10.1$}} \\
  & & & 0.050 
    & 0.047 & 0.17 & 
    & 0.049 & 0.18 & 
    & 0.051 & 0.29 & 
    & 0.050 & 0.14 & 
    & 0.045 & 0.23 & 
    & 0.048 & 13.81 & \\
\cmidrule(lr){2-22}
  & \multirow{2}{*}{\textcolor{black}{1000}}
    & \multirow{2}{*}{10}
    & 0.100 
    & 0.098 & 0.17 & \multicolumn{1}{c}{\multirow{2}{*}{$2.4$}} 
    & 0.097 & 0.13 & \multicolumn{1}{c}{\multirow{2}{*}{$2.0$}} 
    & 0.099 & 0.43 & \multicolumn{1}{c}{\multirow{2}{*}{$5.6$}} 
    & 0.098 & 0.12 & \multicolumn{1}{c}{\multirow{2}{*}{$1.9$}} 
    & 0.096 & 0.30 & \multicolumn{1}{c}{\multirow{2}{*}{$3.8$}} 
    & 0.102 & 15.49 & \multicolumn{1}{c}{\multirow{2}{*}{$15.4$}} \\
  & & & 0.050 
    & 0.047 & 0.22 &  
    & 0.046 & 0.18 & 
    & 0.049 & 2.55 &  
    & 0.048 & 0.17 & 
    & 0.045 & 0.46 &  
    & 0.052 & 29.50 & \\
\cmidrule(lr){2-22}
  & \multirow{2}{*}{\textcolor{black}{1000}}
    & \multirow{2}{*}{5}
    & 0.100 
    & 0.093 & 0.21 & \multicolumn{1}{c}{\multirow{2}{*}{$2.7$}} 
    & 0.096 & 0.15 & \multicolumn{1}{c}{\multirow{2}{*}{$2.1$}} 
    & 0.099 & 1.09 & \multicolumn{1}{c}{\multirow{2}{*}{$8.0$}} 
    & 0.097 & 0.15 & \multicolumn{1}{c}{\multirow{2}{*}{$2.0$}} 
    & 0.095 & 0.78 & \multicolumn{1}{c}{\multirow{2}{*}{$6.7$}} 
    & 0.100 & 20.52 & \multicolumn{1}{c}{\multirow{2}{*}{$23.1$}} \\
  & & & 0.050 
    & 0.047 & 0.28 &  
    & 0.047 & 0.21 & 
    & 0.048 & 12.98 & 
    & 0.046 & 0.21 & 
    & 0.046 & 3.72 &  
    & 0.052 & 47.87 &  \\

\bottomrule
\end{tabular}
\end{adjustbox}
\vspace{-0.25cm}
\end{table*}

\begin{table}
\centering
\scriptsize
\color{black}
\caption{Comparison of CRC-SSL-N with the~\ac{CP}-based multi-source~\ac{DOA} approach for~$K=2$.}
\label{tab:CRC_known_source_compare_with_khurjekar}
\setlength{\tabcolsep}{2.5pt}
\renewcommand{\arraystretch}{1.35}
\sisetup{table-number-alignment = center}

\begin{tabular}{@{}
l
c
c
S[table-format=1.3] S[table-format=2.2] c
S[table-format=1.3] S[table-format=2.2] c
@{}}
\toprule
\multirow{3}{*}{Method}
& \multirow{3}{*}{\makecell{$T_{60}$\\$\mathrm{[ms]}$}}
& \multirow{3}{*}{$\alpha_{\mathrm{MC}}$}
& \multicolumn{3}{c}{Source 1}
& \multicolumn{3}{c}{Source 2} \\
\cmidrule(lr){4-6} \cmidrule(lr){7-9}
&
&
& {\ac{MC}} 
& {PA\si{\percent}}
& \multicolumn{1}{c}{\makecell{MAE\si{\degree}}}
& {\ac{MC}} 
& {PA\si{\percent}}
& \multicolumn{1}{c}{\makecell{MAE\si{\degree}}} \\
\midrule

\multirow{4}{*}{CRC-SSL-N}
& \multirow{2}{*}{400} & 0.100 & 0.092 & 0.09 & \multicolumn{1}{c}{\multirow{2}{*}{$1.6$}} & 0.101 & 0.15 & \multicolumn{1}{c}{\multirow{2}{*}{$2.7$}} \\
&                      & 0.050 & 0.044 & 0.12 &  & 0.051 & 0.19 &  \\
\cmidrule(lr){2-9}
& \multirow{2}{*}{700} & 0.100 & 0.093 & 0.12 & \multicolumn{1}{c}{\multirow{2}{*}{$1.7$}} & 0.098 & 0.19 & \multicolumn{1}{c}{\multirow{2}{*}{$3.0$}} \\
&                      & 0.050 & 0.049 & 0.18 &  & 0.051 & 0.29 &  \\

\midrule

\multirow{4}{*}{\makecell[l]{MDN-CP~\cite{khurjekar2024multi}}}
& \multirow{2}{*}{400} & 0.100 & 0.095 & 2.44 & \multicolumn{1}{c}{\multirow{2}{*}{$8.8$}} & 0.102 & 2.37 & \multicolumn{1}{c}{\multirow{2}{*}{$9.4$}} \\
&                      & 0.050 & 0.051 & 3.55 &  & 0.053 & 4.14 &  \\
\cmidrule(lr){2-9}
& \multirow{2}{*}{700} & 0.100 & 0.097 & 2.52 & \multicolumn{1}{c}{\multirow{2}{*}{$8.7$}} & 0.101 & 3.07 & \multicolumn{1}{c}{\multirow{2}{*}{$10.3$}} \\
&                      & 0.050 & 0.048 & 3.34 &  & 0.052 & 4.84 &  \\

\bottomrule
\end{tabular}
\end{table}

\subsubsection{\textbf{Unknown-source count}}
Performance is evaluated at~$\delta\!=\!0.1$ with risk tolerances $\alpha_{\mathrm{MC}}\!=\!\alpha_{\mathrm{MD}}\!=\!0.1$. After calibration, we select, among all feasible solutions in~$\boldsymbol{\Lambda}^*$, the configuration that minimizes the~\ac{FA} objective.
Next, Algorithm~\ref{alg:iterative_ssl} is run on the test set with the selected detection threshold~$\beta_{\mathrm{TH}}=\hat{\beta}$, and prediction regions are constructed for each detected source using Algorithm~\ref{algo:prediction_set} and the selected thresholds~$(\hat{\lambda}_1,\ldots,\hat{\lambda}_{K_\mathrm{max}})$. We report both the empirical average of each objective across trials and the risk violation rates~$P^\textrm{MC}$ and~$P^\textrm{MD}$, defined as the percentage of trials (i.e. different data splits) in which the corresponding risk exceeds its prescribed limit~$\alpha_{\mathrm{MC}}$ and~$\alpha_{\mathrm{MD}}$, respectively, which should remain below $\delta$.\looseness=-1

Table~\ref{tab:unknown_source_count_results} reports the results for both the synthetic and LOCATA datasets. 
For the synthetic data, we achieve simultaneous control of the~\ac{MC} and~\ac{MD} risks, i.e., the proportion of trials in which $\widehat{\mathcal{R}}^{\mathrm{MC}}(\hat{\boldsymbol{\lambda}})>\alpha_{\mathrm{MC}}$ or $\widehat{\mathcal{R}}^{\mathrm{MD}}(\hat{\boldsymbol{\lambda}})>\alpha_{\mathrm{MD}}$ remains below $\delta$. This confirms that the theoretical guarantees of our framework hold across different acoustic scenes and model types. Both methods, however, are conservative relative to the nominal risk targets when compared to~\textcolor{black}{CRC-SSL-N}, as~\textcolor{black}{PT-SSL-U} provides high-probability control over two risks, whereas~\textcolor{black}{CRC-SSL-N} enforces control in expectation over a single risk. Moreover, compared to~\ac{SRP-PHAT}, SRP-DNN produces more distinct peaks, leading to lower~\ac{FA} rates and smaller~\acp{PA}. Similarly to the synthetic dataset, for the LOCATA dataset both~\ac{MD} and~\ac{MC} risks are simultaneously controlled, albeit at more conservative levels, likely due to residual distribution shift effects discussed in Appendix~\ref{appndx:covariate_shift}.\looseness=-1

\textcolor{black}{
Table~\ref{tab:locata_moving_sources} reports the results for~$K_{\mathrm{max}}=2$ moving sources from the LOCATA dataset. Evidently, CRC-SSL-N closely attains the prescribed~\ac{MC} levels, while imposing the stricter target~$\alpha_{\mathrm{MC}}=0.05$ increases the~\ac{PA}, as expected. The second source exhibits a larger~\ac{PA} and a higher~\ac{MAE} than the first, indicating that greater localization uncertainty is reflected in both larger prediction regions and larger localization errors. For unknown~$K$, PT-SSL-U simultaneously controls the~\ac{MC} and~\ac{MD} risks, with both~$P^{\mathrm{MC}}$ and~$P^{\mathrm{MD}}$ remaining below~$\delta$. 
Imposing a more stringent requirement~$\delta=0.05$ further reduces these violation probabilities at the cost of increased~\ac{PA}, while the~\ac{FA} rate remains nearly unchanged.\looseness=-1
}

\subsection{Sensitivity Analysis}
\begin{figure*}[t]
    \centering

    \subfloat{
        \includegraphics[width=0.88\linewidth]{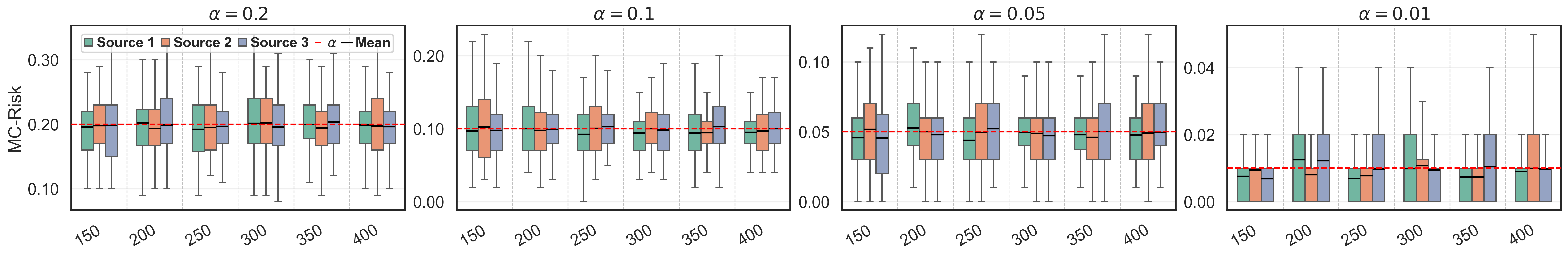}
    }

    \vspace{-0.4cm}

    \subfloat{
        \includegraphics[width=0.88\linewidth]{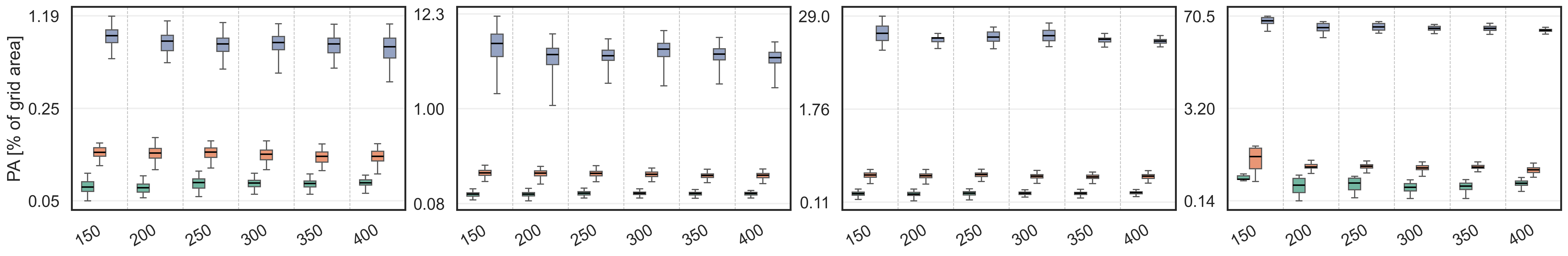}
    }

    \vspace{-0.4cm}

    \subfloat{
        \includegraphics[width=0.88\linewidth]{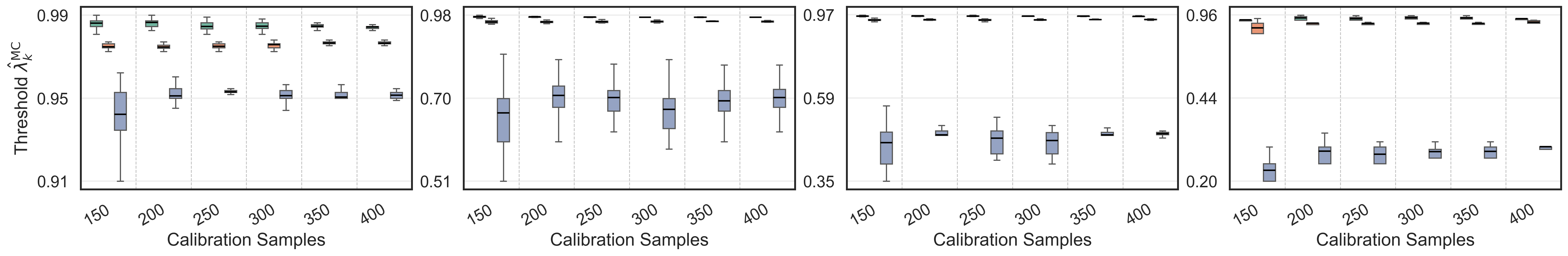}
    }

    \caption{\textcolor{black}{CRC-SSL-N results for varying calibration sizes and $\alpha$ limits, using SRP-DNN with~$T_{60}=\SI{700}{ms}$ and~$\mathrm{SNR}=\SI{15}{dB}$.}}
    \label{fig:ablation_study_known_source_count}
    \vspace{-0.5cm}
\end{figure*}

\begin{figure}
    \centering
    \includegraphics[width=0.83\linewidth]{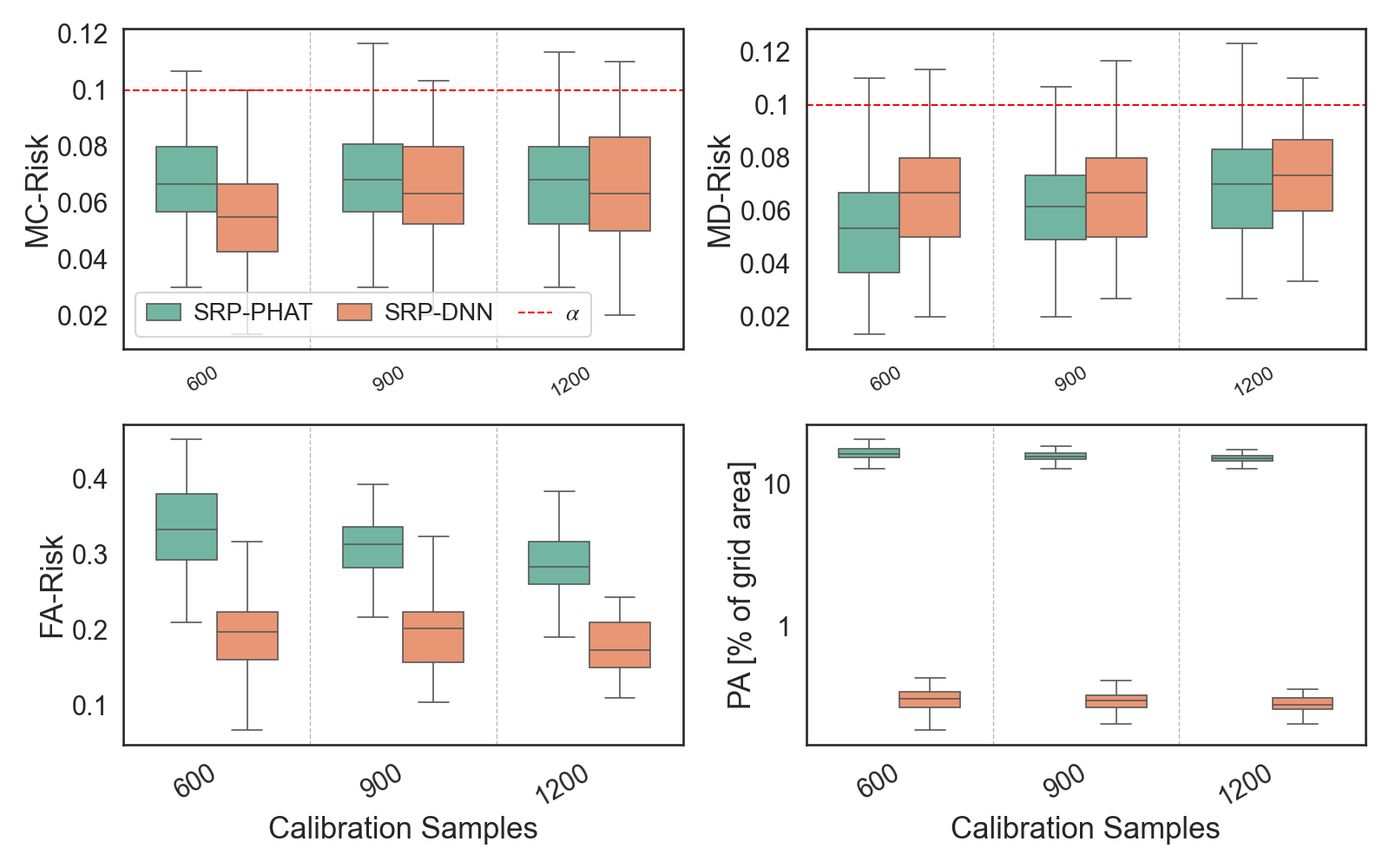}
    \caption{\textcolor{black}{PT-SSL-U results for varying calibration-set sizes, with~$\alpha_{\mathrm{MC}}\!=\!\alpha_{\mathrm{MD}}\!=\!\delta\!=\!0.1$, using SRP-DNN at~$T_{60}\!=\!400$~ms.}}
    \label{fig:calibration_set_size}
\end{figure}

\BL{Since large calibration sets are often unavailable, we examine the sensitivity of the proposed methods to calibration size. For CRC-SSL-N, we consider~$n_{\mathrm{cal}}=\{150, 200,\ldots, 400\}$, under prescribed levels~$\alpha\!\in\!\{0.2,0.1,0.05,0.01\}$.
Fig.~\ref{fig:ablation_study_known_source_count} reports the resulting \ac{MC}, \ac{PA}, and selected threshold~$\hat{\lambda}^{\mathrm{MC}}$.
Across all values of~$\alpha$, the empirical~\ac{MC} remains close to the prescribed target for all~$n_{\mathrm{cal}}$, with stable average~\ac{PA}. Smaller~$n_{\mathrm{cal}}$ increases the variability of the selected thresholds and~\ac{PA}, but does not introduce systematic bias, demonstrating robustness to calibration size. For PT-SSL-U,
we consider~$n_{\mathrm{cal}}\!\in\!\{600,900,1200\}$ and $\alpha_{\mathrm{MC}}=\alpha_{\mathrm{MD}}=\delta=0.1$.
Fig.~\ref{fig:calibration_set_size} summarizes the results. For all~$n_{\mathrm{cal}}$, both~\ac{MC} and~\ac{MD} risks remain well controlled. Smaller $n_{\mathrm{cal}}$ yield more conservative behavior, with fewer trials exceeding the risk thresholds. As $n_{\mathrm{cal}}$ decreases, the mean~\ac{FA} and~\ac{PA} increase slightly and their distributions become more dispersed. Overall, we conclude that using smaller calibration sets is acceptable.\looseness=-1
}

\BL{
We conclude this section with a remark on~$K_{\max}$.
In our experiments,~$K_{\max}$ is set to the maximum ground-truth source count. However, the validity of PT-SSL-U does not require this upper bound to be tight. An overly loose~$K_{\max}$ directly affects computational load, since the configuration-space size scales as~$|\boldsymbol{\Lambda}_g|=N^{K_{\max}+1}$. In addition, if~$K_{\max}$ is much larger than the true source count, Algorithm~\ref{alg:iterative_ssl} may generate additional likelihood maps with dominant spurious peaks. PT-SSL-U is designed to control this behavior through the detection threshold~$\beta_{\mathrm{TH}}$ and the~\ac{FA} objective. Increasing~$\beta_{\mathrm{TH}}$ reduces~\acp{FA} but may increase~\acp{MD}, and vice versa. Consequently, the added~\ac{FA}/\ac{MD} trade-off may result in more conservative settings.
}
\looseness=-1

\subsection{Comparison with~\ac{UQ} Baselines}
\label{sec:comp_with_baselines}

\BL{In this section, we evaluate CRC-SSL-N against two categories of~\ac{UQ} approaches for multi-speaker localization. The first category is a direct comparison to a~\ac{CP}-based localization approach~\cite{khurjekar2024multi}, which, like ours, provides set-valued localization outputs with coverage guarantees. The second category is an indirect comparison to uncertainty proxies derived directly from the spatial likelihood map. These proxies do not provide formal coverage guarantees, but are commonly used to assess the confidence of localization estimates.\looseness=-1}

\BL{
\textbf{CP-based Comparison.} As a baseline, we consider the method in~\cite{khurjekar2024multi}, based on a~\ac{MDN}~\cite{bishop1994mixture} followed by conformal prediction. The original `MDN-CP' approach was developed for azimuth estimation of narrow-band sources in anechoic environments. 
To adapt it to our setting, we use the likelihood map produced by the~\ac{CRNN} model in Section~\ref{sec:srp-dnn} as an input to a \ac{CNN}, which predicts~$K$ azimuth--elevation pairs (see further details in Appendix~\ref{appendix:cp_for_doa_khurjekar}).
As shown in Table~\ref{tab:CRC_known_source_compare_with_khurjekar}, both methods meet the prescribed~\ac{MC} targets tightly. However,~\cite{khurjekar2024multi} consistently yields larger~\acp{PA} across all tested conditions. This demonstrates the improved efficiency of our flexible region construction procedure.}\looseness=-1

\begin{figure}
    \centering
    \includegraphics[width=0.8\linewidth]{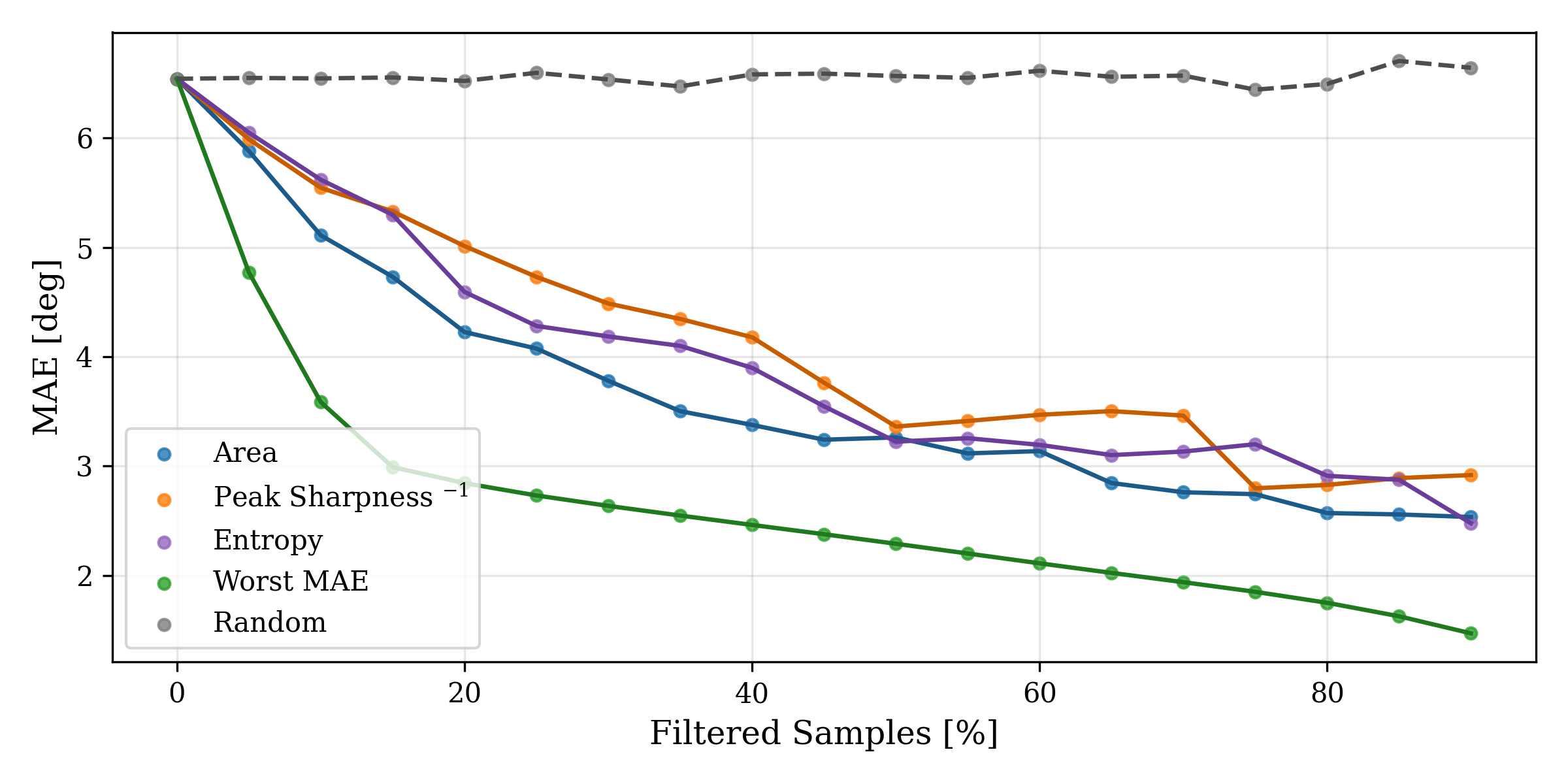}
    \caption{
        \textcolor{black}{Comparison with~\ac{UQ} baselines.}
    }
    \label{fig:uncertainty_methods}
    \vspace{-0.8em}
\end{figure}

\BL{
\textbf{Confidence Assessment.} We treat the prediction region produced by CRC-SSL-N as a proxy for uncertainty and compare it to peak sharpness and map entropy. The peak sharpness is defined as a local Laplacian contrast around the detected peak.
For this analysis, we define the angular error between the true and estimated~\ac{DOA} as~$e(\mathbf{p}^*, \hat{\mathbf{p}}) = \arccos\!\left( \mathbf{u}(\mathbf{p}^*)^{\mathsf{T}} \mathbf{u}(\hat{\mathbf{p}}) \right)$, 
where~$\mathbf{u}(\cdot)$ is the direction vector, defined in~\eqref{eq:tdoa}. 
We compute the~\ac{MAE} by averaging~$e(\cdot)$ over all test samples.
For each uncertainty measure, samples are ranked by the measure and the most uncertain fraction is progressively removed. The~\ac{MAE} is then computed on the remaining samples after each rejection step. This procedure assesses how well each uncertainty measure aligns with localization error: a more informative measure should lead to a stronger reduction in~\ac{MAE} as uncertain samples are rejected. As references, we additionally report an oracle rejection strategy, which removes samples with the largest~\ac{MAE}, and a random rejection strategy. Fig.~\ref{fig:uncertainty_methods} reports the results for SRP-DNN with~$T_{60}=700\,\mathrm{ms}$, $\mathrm{SNR}=15\,\mathrm{dB}$ and~$K=3$ sources. We observe that our method leads to greater~\ac{MAE} reduction, indicating that it better quantifies estimation uncertainty. \looseness=-1 }  

\subsection{Application to Beamforming}
\label{sec:application}
\BL{
We exemplify the utilization of our~\ac{UQ} framework for improving speech enhancement capabilities. We design a beamformer that exploits the prediction regions $\mathcal{C}_{\mathrm{tar}}=\left\{\mathbf{p}_i\right\}_{i=1}^{N_{\mathrm{tar}}}$ and~$\mathcal{C}_{\mathrm{int}}=\left\{\mathbf{p}_j\right\}_{j=1}^{N_{\mathrm{int}}}$ for a target and an interfering source, respectively. We employ the~\ac{LCMV}~\cite{lcmv} beamformer, which minimizes the output power while enforcing a set of linear constraints corresponding to the target and interfering directions. For each~$\mathbf{p}$ in either prediction region, we evaluate the steering vector~$\mathbf{a}(\mathbf{p},f)=
[1, \mathrm{IPD}_{21}(f;\mathbf{p}),\ldots,\mathrm{IPD}_{M1}(f;\mathbf{p})]^{\mathsf T}$,
where~$\mathrm{IPD}_{mm'}(\cdot)$ is defined in~\eqref{eq:IPD}. The constraint matrix~$\mathbf{C}(f)\in\mathbb{C}^{M\times L}$ is then formed by stacking the steering vectors associated with all constrained directions, where $L=N_{\mathrm{tar}}+N_{\mathrm{int}}$. If $L>M$, we simply select representative positions from each region.
The beamformer weights are given by:\looseness=-1
\begin{equation}
\mathbf{w}_{\mathrm{LCMV}}(f)=\mathbf{R}_{xx}(f)^{-1}\mathbf{C}(f)
\left(\mathbf{C}(f)^{H}\mathbf{R}_{xx}(f)^{-1}\mathbf{C}(f)\right)^{-1}
\mathbf{g}.\notag
\end{equation}
where~$\mathbf{R}_{xx}(f)\in\mathbb{C}^{M\times M}$
denotes the spatial covariance matrix of the measured signal defined at~\eqref{eq:problem_formulation_in_freq} and~$\mathbf{g}\!=\!\left[\mathbf{1}_{N_{\mathrm{tar}}},\mathbf{0}_{N_{\mathrm{int}}}\right]^{\mathsf T}$, where~$\mathbf{1}_{N}$ and~$\mathbf{0}_{N}$ denote length-$N$ all-one and all-zero vectors, respectively. The vector~$\mathbf{g}$ enforces unit gain for directions associated with the target prediction region while placing nulls in the directions belonging to the interferer prediction region.
}\looseness=-1

\BL{
Experiments are conducted on a synthetic dataset with~$K\!=\!2$, $T_{60}\!=\!\SI{250}{ms}$ and $\mathrm{SNR}=\SI{20}{dB}$, using the same room setup used in the previous experiments. Each source serves once as the target and once as the interferer. We use~$400$ calibration and~$100$ test samples, repeating the experiment over~$50$ random data partitions. We apply the~\ac{LCMV} beamformer as a downstream action using three~\ac{SSL} inputs: (i) oracle~\acp{DOA}, (ii) CRC-SSL-N prediction regions calibrated with~$\alpha=0.1$, and (iii)~\ac{PD}. Table~\ref{tab:lcmv_exp} summarizes the performance, assessed using~\ac{SI-SDR} and~\ac{ESTOI}. The proposed CRC-SSL-N framework achieves near-oracle~\ac{DOA} performance while outperforming the~\ac{PD} baseline, demonstrating the utility of the constructed prediction regions for beamforming.
}\looseness=-1

\begin{table}[t]
\scriptsize
\centering
\color{black}
\caption{LCMV beamformer evaluation results for~$K=2$.}
\begin{tabular}{llcccc}
\hline
\noalign{\vskip 2pt}
Metric & Source & Noisy & Oracle &~\ac{PD} & CRC-SSL-N \\
\noalign{\vskip 2pt}
\hline
\multirow{2}{*}{\makecell[l]{SI-SDR~{[dB]}}}
 & $k=1$ & -2.977 & 3.964 & 3.756 & 3.932 \\
 & $k=2$ & -2.883 & 3.914 & 3.587 & 3.777 \\
\midrule
\multirow{2}{*}{ESTOI}
 & $k=1$ & 0.514 & 0.763 & 0.762 & 0.762 \\
 & $k=2$ & 0.509 & 0.774 & 0.772 & 0.773 \\
\hline
\label{tab:lcmv_exp}
\vspace{-0.25cm}
\end{tabular}
\end{table}

\begin{table*}[t]
\centering
\caption{PT-SSL-U results, with \ac{MC} and \ac{MD} controlled at~$\alpha_{\mathrm{MC}}=\alpha_{\mathrm{MD}}=0.1$ with significance level $\delta = 0.1$.} 
\label{tab:unknown_source_count_results}

\scriptsize
\setlength{\tabcolsep}{3.5pt}
\providecommand{\NA}{\multicolumn{1}{c}{--}}

\begin{tabular}{@{} c c c *{14}{S[table-format=2.2]} @{}}
\toprule
\multirow{4}{*}{Dataset} &
\multirow{4}{*}{$K_{\mathrm{max}}$} &
\multirow{4}{*}{\makecell[c]{$T_{60}$\\{[ms]}}}
& \multicolumn{7}{c}{SRP-PHAT}
& \multicolumn{7}{c}{SRP-DNN} \\

\cmidrule(lr){4-10} \cmidrule(lr){11-17}
& & 
& {$P^{\mathrm{MC}}$} & {MC} & {$P^{\mathrm{MD}}$} & {MD} & {FA} & {PA\%} & {\makecell{MAE\si{\degree}}}
& {$P^{\mathrm{MC}}$} & {MC} & {$P^{\mathrm{MD}}$} & {MD} & {FA} & {PA\%} & {\makecell{MAE\si{\degree}}} \\
\midrule

\multirow{2}{*}{Synthetic}
& \multirow{2}{*}{3}
& 400
& 0.06 & 0.06 & 0.09 & 0.06 & 0.31 & 16.58 & \multicolumn{1}{c}{$8.7$}
& 0.04 & 0.06 & 0.08 & 0.06 & 0.21 & 0.44 & \multicolumn{1}{c}{$2.7$} \\

& & 700
& 0.07 & 0.07 & 0.04 & 0.06 & 0.35 & 21.53 &  \multicolumn{1}{c}{$10.9$}
& 0.03 & 0.06 & 0.06 & 0.07 & 0.29 & 0.58 & \multicolumn{1}{c}{$3.0$} \\

\midrule

LOCATA
& \multirow{2}{*}{2}
& \multirow{2}{*}{550}
& 0.05 & 0.05 & 0.03 & 0.06 & 0.22 & 9.80 & \multicolumn{1}{c}{$6.6$}
& 0.03 & 0.04 & 0.04 & 0.05 & 0.16 & 0.37 &  \multicolumn{1}{c}{$2.4$} \\

Synthetic
& & 
& 0.04 & 0.05 & 0.04 & 0.05 & 0.30 & 6.70 & \multicolumn{1}{c}{$5.6$}
& 0.02 & 0.04 & 0.04 & 0.05 & 0.11 & 0.30 & \multicolumn{1}{c}{$2.5$} \\

\bottomrule
\end{tabular}
\vspace{-0.25cm}
\end{table*}


\begin{table*}[t]
\centering

\color{black}
\caption{Moving-source LOCATA using the SRP-DNN backend with $K_{\mathrm{max}}=2$.}
\label{tab:locata_moving_sources}

\scriptsize
\setlength{\tabcolsep}{3.5pt}
\renewcommand{\arraystretch}{1.2}
\sisetup{table-number-alignment = center}
\providecommand{\NA}{\multicolumn{1}{c}{--}}

\begin{tabular}{@{} c *{9}{S[table-format=1.3]} *{10}{S[table-format=2.2]} @{}}
\toprule

\multicolumn{10}{c}{CRC-SSL-N} 
& \multicolumn{10}{c}{PT-SSL-U} \\
\cmidrule(r){2-10} \cmidrule(l){11-20}

&
\multicolumn{3}{c}{$K=1$}
& \multicolumn{6}{c}{$K=2$}
& \multicolumn{10}{c}{\multirow{3}{*}{Unknown $K$}} \\
\cmidrule(lr){2-4} \cmidrule(lr){5-10}

& \multicolumn{3}{c}{Source 1}
& \multicolumn{3}{c}{Source 1}
& \multicolumn{3}{c}{Source 2} \\
\cmidrule(lr){2-4} \cmidrule(lr){5-7} \cmidrule(lr){8-10}

{$\alpha_{\mathrm{MC}}$}
& {\ac{MC}} & {PA\%} & {\makecell{MAE\si{\degree}}}
& {\ac{MC}} & {PA\%} & {\makecell{MAE\si{\degree}}}
& {\ac{MC}} & {PA\%} & {\makecell{MAE\si{\degree}}}
& {$\delta$}
& {$\alpha_{\mathrm{MC}}$}
& {$\alpha_{\mathrm{MD}}$}
& {$P^{\mathrm{MC}}$}
& {MC}
& {$P^{\mathrm{MD}}$}
& {MD}
& {FA}
& {PA\%}
& {\makecell{MAE\si{\degree}}} \\
\cmidrule(r){1-10}\cmidrule(l){11-20}

0.100 & 0.101 & 1.75 & \multicolumn{1}{c}{\multirow{2}{*}{$7.7$}}
& 0.100 & 1.52 & \multicolumn{1}{c}{\multirow{2}{*}{$6.8$}}
& 0.098 & 5.51 & \multicolumn{1}{c}{\multirow{2}{*}{$12.0$}}
& 0.100 & 0.100 & 0.100 & 0.080 & 0.057 & 0.050 & 0.056 & 0.28 & 6.21 & \multicolumn{1}{c}{\multirow{2}{*}{$8.6$}} \\

0.050 & 0.047 & 2.50 & 
& 0.950 & 2.22 & 
& 0.049 & 8.89 & 
& 0.050 & 0.100 & 0.100 & 0.030 & 0.046 & 0.030 & 0.056 & 0.29 & 6.60 &  \\

\bottomrule
\end{tabular}
\end{table*}


\section{Limitations and Future Work}
\label{sec:limiations_future_work}
\textcolor{black}{
Overall, the proposed frameworks provide a practical, statistically principled means of calibrating reliability in~\ac{SSL} systems.
Their main limitations are the exchangeability assumption and the need for large calibration datasets. Exchangeability between the calibration and test data is fundamental to the theoretical guarantees of~\ac{CP} but may be violated by changes in reverberation, noise level, or microphone-array configuration, making representative calibration data difficult to obtain. Moreover, Pareto-Testing itself requires substantial calibration data, since the available samples must be divided into separate optimization and hypothesis-testing sets. Future work should investigate adaptive or distribution-shift-aware calibration strategies~\cite{cp_cov_shift}, as well as more data-efficient~\ac{MHT} procedures that mitigate the cost of data splitting while preserving statistical guarantees~\cite{einbinder2025semi}.}\looseness=-1
\section{Conclusion}
\label{sec:conclusion}
We present two frameworks that provide finite-sample guarantees for~\ac{UQ} in~\ac{SSL}. For the known-source-count setting, we employ \textcolor{black}{CRC-SSL-N} to construct prediction regions with controlled~\ac{MC}. For the unknown-source-count setting, we use \textcolor{black}{PT-SSL-U} to obtain a statistically efficient risk-controlling configuration selection with simultaneous control of~\ac{MC} and~\ac{MD}, while encouraging minimal~\ac{FA} and the smallest~\acp{PA} among admissible configurations. Experiments on both synthetic and real-world recordings demonstrate that these guarantees are maintained across diverse acoustic conditions and models.\looseness=-1 
\appendices  
\section{Mathematical Details}
\subsection{Computation of Valid $p$-Values}
\label{appendix:p_values}
A convenient way to compute $p$-values is through concentration inequalities, but their tightness depends on the underlying loss function. For binary loss functions (e.g.,~\ac{MC} defined in~\eqref{eq:mis_cov_loss_averaged}), the Binomial tail bound is tight. For loss function bounded in $[0, 1]$, the~\ac{HB} bound~\cite{bates2021distribution} provides finite-sample guarantees with strong empirical effectiveness, but for more general loss functions it yields valid $p$-values only asymptotically. Moreover, \ac{HB} does not adapt to the (unknown) loss variance, leading to a loose bound. 
The \ac{WSR} bound~\cite{wsr}, originally developed for random variables supported on $[0,1]$, adapts to the empirical variance and can be seamlessly extended to arbitrary finite intervals $[A,B]$~\cite{einbinder2025semi}.\looseness=-1

For completeness, we outline the derivation of valid~$p$-values based on the~\ac{WSR} bound. 
Let~\(L_i({\boldsymbol{\lambda}})\!\in\![0,1]\), \(i=1,\ldots,n\), be i.i.d.\ random variables, and consider testing the null hypothesis~\(\mathcal{R}({\boldsymbol{\lambda}})\!>\!\alpha\). 
Under the paradigm of \emph{testing-by-betting}~\cite{ramdas_Hypothesis_Testing_with_e_values}, sequential bets are placed against the null, and substantial growth of the resulting capital serves as evidence against it. 
Define the capital process, initialized at unit wealth, by\looseness=-1
\begin{equation}
    \mathcal{K}_i(\boldsymbol{\lambda}, \alpha)
    = \prod_{t=1}^{i} \left\{ 1 - \nu_t\,
    \bigl(L_t({\boldsymbol{\lambda}}) - \alpha\bigr) \right\}, 
    \;\; i=1,\ldots,n,
\end{equation}
where $\nu_t$ denotes the betting strategy, which depends on past loss observations $L_1({\boldsymbol{\lambda}}), \ldots, L_{t-1}({\boldsymbol{\lambda}})$ and error significance $\delta$. 
Under $\mathcal{H}_0$, the process $\{\mathcal{K}_i\}$ is a nonnegative supermartingale with $\mathbb{E}[\mathcal{K}_i]\le 1$.
By Ville's inequality~\cite{doob1939jean} it follows that
\begin{equation}
    \mathbb{P}\left (\max_{i=1,\ldots,n} \;\mathcal{K}_i(\boldsymbol{\lambda}, \alpha)\ge \frac{1}{\delta}\right)\le\delta.
\label{eq:ville_theorem}
\end{equation}
Thus, under~$\mathcal{H}_0$ and for any betting strategy, the probability that the capital process ever exceeds~$1/\delta$ is at most $\delta$. Thus, observing~$\mathcal{K}_i(\boldsymbol{\lambda},\alpha)\!\ge\! 1/\delta$ for some~$i$ constitutes strong evidence against $\mathcal{H}_0$ and justifies its rejection at level~$\delta$.
Since Ville's inequality in~\eqref{eq:ville_theorem} can be expressed as~$\mathbb{P}\left (\min_{i} \;\frac{1}{\mathcal{K}_i(\boldsymbol{\lambda}, \alpha)}\le \delta\right)\le\delta$, the corresponding~$p$-value is
\begin{equation}
    p(\boldsymbol{\lambda}, \alpha)=\min_{i=1, \ldots, n} \;\;\frac{1}{\mathcal{K}_i(\boldsymbol{\lambda}, \alpha)}.
    \label{eq:wsr_pvalue}
\end{equation}
Algorithm~\ref{algo:wsr_pvalue} summarizes~\ac{WSR} $p$-value computation~\cite{einbinder2025semi}.
\looseness=-1

\begin{algorithm}[!t]
\caption{WSR $P$-Value Evaluation}
\label{algo:wsr_pvalue}
\begin{algorithmic}[1]
\Require
Collection of $n$ i.i.d.\ random variables $L_i \in [A,B]$; error significance $\delta$; risk tolerance $\alpha$.
\Ensure Valid $p$-value for the null hypothesis $\mathcal{H}:\widehat{\mathcal{R}}_n>\alpha$
\For{$i = 1,\ldots,n$}
    \State $\displaystyle
    \hat{\mu}_i \gets
     \frac{\frac{1}{2}+\sum_{j=1}^{i} L_j}{1+i},\,\, \hat{\sigma}^2_i \gets
     \frac{\frac{1}{4}+\sum_{j=1}^{i} ( L_t - \hat{\mu}_t )^{2}}{1+i}$ 
    \State $\displaystyle
    \nu_i \gets \min \!\left\{
        \tfrac{1}{B-A},\;
        \sqrt{ \tfrac{2\log(1/\delta)}{\,n\hat{\sigma}^2_{i - 1}\,} }
      \right\}$
    \State $\displaystyle
    \mathcal{K}_i \gets
    \prod_{t=1}^{i}\bigl\{1 - \nu_t ( L_t - \alpha )\bigr\}$
\EndFor
\State $p\gets\min_{i=1,\ldots, n} \;\;\frac{1}{\mathcal{K}_i}.$
\State \Return $p$
\end{algorithmic}
\end{algorithm}
\vspace{-3pt}
\subsection{Recovering the Pareto Optimal Front}
\label{appendix:pareto_opt_set}

Algorithm~\ref{alg:recover_pareto} describes the filtering of the Pareto-optimal set.\looseness=-1
\begin{algorithm}[t]
\caption{Recover Pareto-Optimal Set}
\label{alg:recover_pareto}
\begin{algorithmic}[1]
\Require Discrete space $\boldsymbol{\Lambda}_g$; Objective functions $\{\mathcal{R}_1,\ldots,\mathcal{R}_r\}$
\Ensure Pareto-optimal set $\boldsymbol{\Lambda}_{\mathrm{par}}$
    \State $\boldsymbol{\Lambda}_o \gets \boldsymbol{\Lambda}_g$
    \For{$\boldsymbol{\lambda} \in \boldsymbol{\Lambda}_g$}
        \If{$\exists \boldsymbol{\lambda}' \in \boldsymbol{\Lambda}_g\setminus\{\boldsymbol{\lambda}\}$ s.t. \\
            \hspace{\algorithmicindent} $\forall i\!:\!\mathcal{R}_i(\boldsymbol{\lambda}') \le \mathcal{R}_i(\boldsymbol{\lambda})$ 
             \textbf{and} $\exists j\!:\!\mathcal{R}_j(\boldsymbol{\lambda}') < \mathcal{R}_j(\boldsymbol{\lambda})$}
            \State $\boldsymbol{\Lambda}_o \gets \boldsymbol{\Lambda}_o \setminus \{\boldsymbol{\lambda}\}$
        \EndIf
    \EndFor
    \State $\boldsymbol{\Lambda}_{\mathrm{par}} \gets \boldsymbol{\Lambda}_o$
    \State \Return $\boldsymbol{\Lambda}_{\mathrm{par}}$
\end{algorithmic}
\end{algorithm}

\section{Additional Experimental Details \& Results}
\label{sec:app_b}

\subsection{Mitigating Simulation–Real Distribution Shifts}
\label{appndx:covariate_shift}
\begin{figure}[t]
\vspace{-1em}
\centering
\subfloat[$K=1,k=1$]{%
\includegraphics[width=0.4\columnwidth]{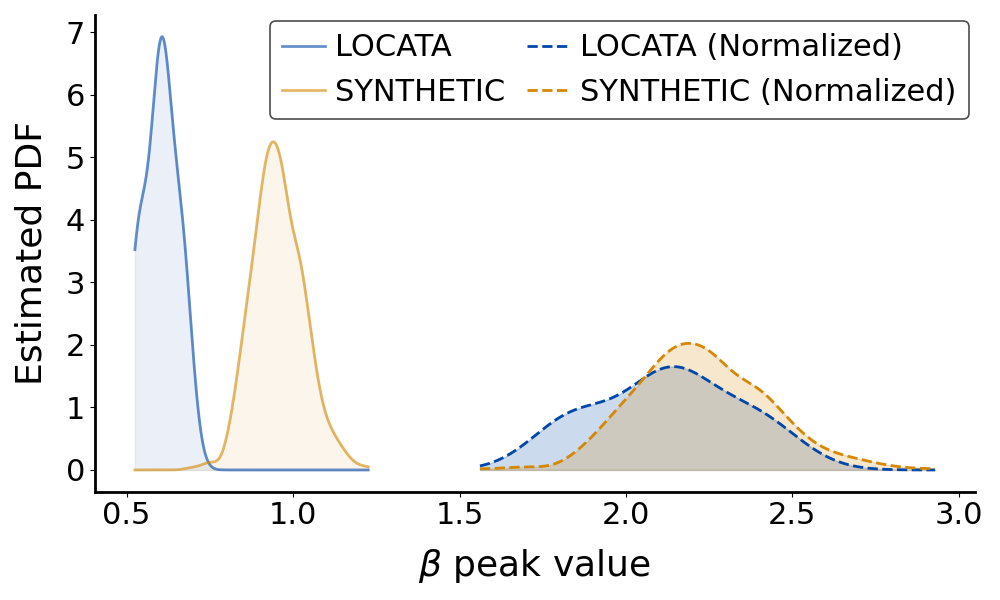}}%
\hspace{0.1\columnwidth}%
\subfloat[$K\!=\!1,k\!=\!2\;$(noise)]{%
\includegraphics[width=0.4\columnwidth]{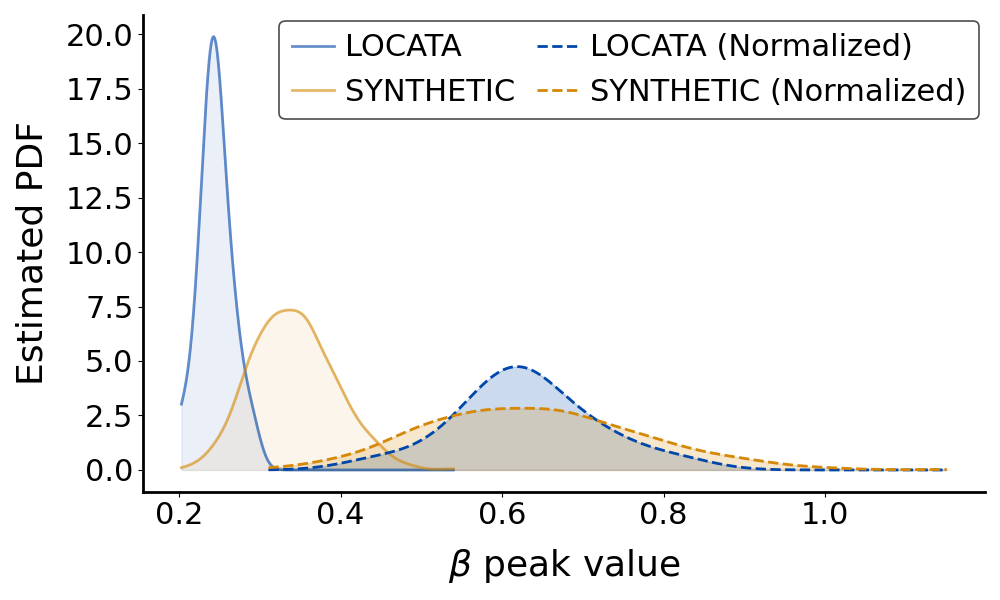}}\\
\vspace{-1em}
\subfloat[$K=2,k=1$]{%
\includegraphics[width=0.4\columnwidth]{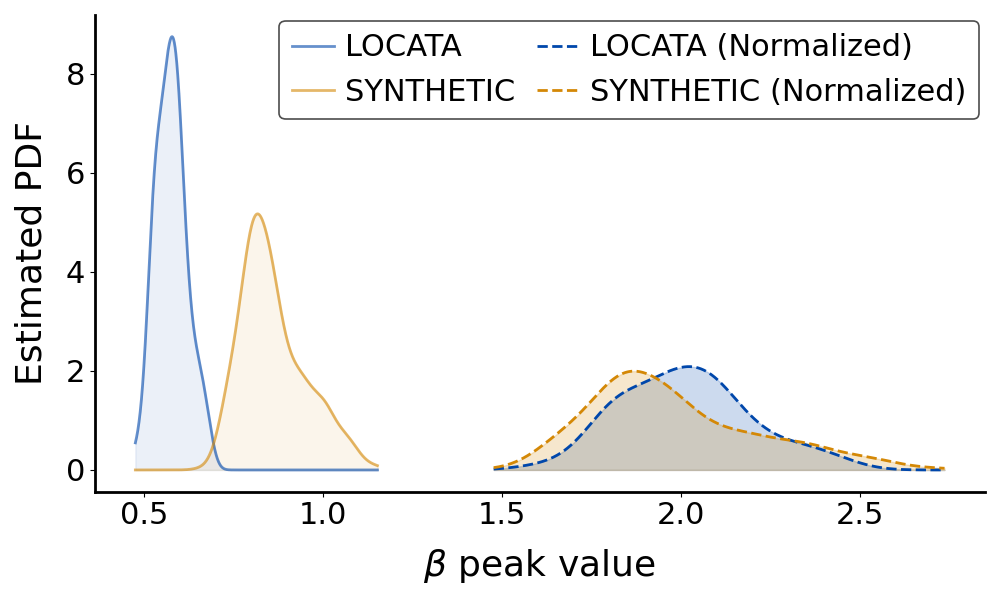}}%
\hspace{0.1\columnwidth}%
\subfloat[$K=2,k=2$]{%
\includegraphics[width=0.4\columnwidth]{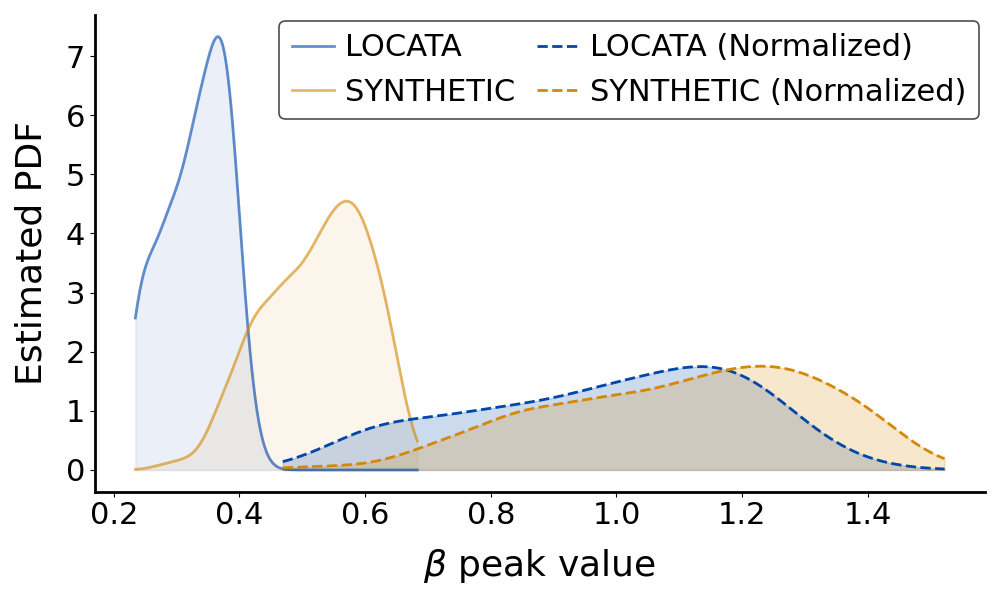}}
\caption{Estimated PDFs of $\beta$ peak values for $K_{\mathrm{max}}=2$.}
\label{fig:cov_shift}
\vspace{-5pt}
\end{figure}

Despite closely matching the acoustic conditions of the LOCATA-matched synthetic dataset used for calibration with those of the real-world LOCATA recordings used for testing, a noticeable distribution shift persists in the detected peak statistics, as shown in Fig.~\ref{fig:cov_shift}. Although non-exchangeability remains, we partially mitigate this shift using an affine transformation that preserves the spatial structure of the likelihood map. Specifically, the $i$-th calibration map is normalized as\looseness=-1
\begin{equation}
\label{eq:transformation}
\tilde{\Phi}_i^{(k)} =
\frac{
\Phi_i^{(k)} - q_{0.5}\!\left(\mathcal{D}^{\Phi}_{\mathrm{cal}}\right)
}{
q_{0.99}\!\left(\mathcal{D}^{\Phi}_{\mathrm{cal}}\right)
-
q_{0.5}\!\left(\mathcal{D}^{\Phi}_{\mathrm{cal}}\right)
}.
\end{equation}
where~$q_{\tau}(\cdot)$ denotes the $\tau$-quantile
and~$\mathcal{D}^{\Phi}_{\mathrm{cal}}
= \{\{\Phi_i^{(k)}\}_{k=1}^{K_{\max}}\}_{i\in\mathcal{D}_\mathrm{cal}}$. An analogous normalization is applied to the test likelihood maps using the test set statistics. As illustrated in Fig.~\ref{fig:cov_shift}, this normalization substantially improves alignment between the calibration and test distributions.\looseness=-1

\subsection{\ac{MC} Risk Control Across Multiple Rooms}
\label{appndx:multi_room_crc}
Controlling the \ac{MC} risk simultaneously across multiple rooms is important for practical deployment, as acoustic conditions vary and per-room recalibration is often impractical. To allow robustness to such variability, we employ unified calibration and test sets spanning multiple room conditions.
Specifically, the calibration and test sets comprise three simulated rooms with fixed dimensions and constant \ac{SNR}, while varying the reverberation time $T_{60}\in\{400,550,700\}$ ms. The results are reported in Table~\ref{tab:multilpe_rooms}.
As shown, the empirical coverage remains very close to the nominal level across all source counts, indicating effective control of~\ac{MC} risk under heterogeneous acoustic conditions. The~\ac{PA} falls reasonably between the values obtained for $T_{60}=400$ ms and $T_{60}=700$ ms reported in Table~\ref{tab:CRC_known_source_merged}. Note, however, that this approach provides only marginal coverage guarantees over the mixture distribution of rooms, whereas room-specific calibration yields coverage guarantees tailored to each setup individually.

\begin{table}[t]
\centering
\caption{CRC-SSL-N controlled simultaneously across multiple rooms ($T_{60}\!\in\!\{400,550,700\}$\,ms, \ac{SNR}$=15$\,dB).\looseness=-1}
\label{tab:multilpe_rooms}
\scriptsize
\setlength{\tabcolsep}{3.2pt}
\renewcommand{\arraystretch}{1.1}
\sisetup{table-number-alignment = center}

\resizebox{\columnwidth}{!}{%
\begin{tabular}{@{} c c *{12}{S[table-format=1.3]} @{}}
\toprule
\multirow{4}{*}{}
  & \multirow{4}{*}{$\alpha_{\mathrm{MC}}$}
  & \multicolumn{2}{c}{$K=1$}
  & \multicolumn{4}{c}{$K=2$}
  & \multicolumn{6}{c}{$K=3$} \\
\cmidrule(lr){3-4} \cmidrule(lr){5-8} \cmidrule(lr){9-14}

  &  & \multicolumn{2}{c}{Source 1}
       & \multicolumn{2}{c}{Source 1} & \multicolumn{2}{c}{Source 2}
       & \multicolumn{2}{c}{Source 1} & \multicolumn{2}{c}{Source 2} & \multicolumn{2}{c}{Source 3} \\
\cmidrule(lr){3-4} \cmidrule(lr){5-6} \cmidrule(lr){7-8}
\cmidrule(lr){9-10} \cmidrule(lr){11-12} \cmidrule(lr){13-14}

  &  & \multicolumn{1}{c}{\ac{MC}} & \multicolumn{1}{c}{PA\si{\%}}
       & \multicolumn{1}{c}{\ac{MC}} & \multicolumn{1}{c}{PA\si{\%}}
       & \multicolumn{1}{c}{\ac{MC}} & \multicolumn{1}{c}{PA\si{\%}}
       & \multicolumn{1}{c}{\ac{MC}} & \multicolumn{1}{c}{PA\si{\%}}
       & \multicolumn{1}{c}{\ac{MC}} & \multicolumn{1}{c}{PA\si{\%}}
       & \multicolumn{1}{c}{\ac{MC}} & \multicolumn{1}{c}{PA\si{\%}} \\
\midrule

\multirow{2}{*}{\makecell{SRP- \\ PHAT}}
  & 0.100 & 0.098 & 0.18 & 0.099 & 0.18 & 0.101 & 27.15 & 0.095 & 0.19 & 0.101 & 8.26 & 0.097 & 63.68 \\
  & 0.050 & 0.047 & 0.27 & 0.045 & 0.25 & 0.050 & 55.66 & 0.043 & 0.27 & 0.051 & 37.78 & 0.049 & 80.82 \\
  
\midrule

\multirow{2}{*}{\makecell{SRP- \\ DNN}}
  & 0.100 & 0.094 & 0.10 & 0.096 & 0.11 & 0.097 & 0.17 & 0.091 & 0.10 & 0.093 & 0.15 & 0.103 & 2.78 \\
  & 0.050 & 0.051 & 0.14 & 0.046 & 0.14 & 0.045 & 0.25 & 0.042 & 0.13 & 0.046 & 0.20 & 0.047 & 13.64 \\

\bottomrule
\end{tabular}
}
\end{table}

\subsection{Pareto-Testing for a Known Source Count}
\label{appndx:ltt_known_source_count}
We propose an alternative  to the CRC-SSL-N framework of Section~\ref{subsec:known_source_count} using Pareto-Testing, as in the unknown-source-count setting; we refer to this variant as `PT-SSL-N'. The goal is to select a configuration $\boldsymbol{\lambda} = (\lambda_1,\ldots,\lambda_K)$ by solving the following constrained optimization problem\looseness=-1
\begin{equation}
\label{eq:known_source_count_pareto_objective}
\hat{\boldsymbol{\lambda}}
= \arg\min_{\lambda \in \boldsymbol{\Lambda}}
\mathcal{R}^{\mathrm{PA}}(\boldsymbol{\lambda})
\quad \text{s.t.} \quad
\mathbb{P}\!\left(
\mathcal{R}^{\mathrm{MC}}(\boldsymbol{\lambda})
\le \alpha_{\mathrm{MC}}
\right) \ge 1 - \delta, \notag
\end{equation}
where~$\boldsymbol{\Lambda}\!=\!\Lambda_1^{\mathrm{MC}} \times \cdots \times \Lambda_K^{\mathrm{MC}}$.
Here, we aggregate the~\ac{MC} and~\ac{PA} measures across sources, in contrast to~\ac{CRC}, where they are defined per source.
Note also that while both approaches aim to control the~\ac{MC} risk at level $\alpha_{\mathrm{MC}}$, Pareto-Testing provides high-probability guarantees, whereas~\ac{CRC} ensures risk control only in expectation, as formalized in  Theorem~\ref{theo:risk_control}. 
Results for the Pareto-Testing approach are reported in Table~\ref{tab:Pareto_known_srouce_count}. The nominal coverage level is never violated across reverberation times or localization methods. Compared to the~\ac{CRC} results in Table~\ref{tab:CRC_known_source_merged}, Pareto-Testing yields more conservative~\ac{MC} control and slightly larger~\acp{PA}, reflecting the stronger (high-probability) guarantees it enforces.

\subsection{MDN-CP}
\label{appendix:cp_for_doa_khurjekar}
\textcolor{black}{
We adapt the MDN-CP in~\cite{khurjekar2024multi} to our setting for fair comparison. While~\cite{khurjekar2024multi} models only azimuth and uses the multi-channel covariance matrix as input, we jointly model azimuth and elevation, and instead use a likelihood map~$\Phi_i$ as input. 
Our~\ac{DNN} model consists of a lightweight~\ac{CNN} followed by fully connected layers. The output layer maps the extracted features to the parameters of the conditional~\ac{DOA} density expressed as
{
\setlength{\abovedisplayskip}{2pt}
\setlength{\belowdisplayskip}{2pt}
\begin{equation}
p(\mathbf{p}\mid\Phi_i)
=
\sum_{k=1}^{K}
\pi_i^k\,
\mathcal{N}\!\left(
\mathbf{p};
\widehat{\boldsymbol{\mu}}_i^k,
\widehat{\boldsymbol{\Sigma}}_i^k
\right),
\end{equation}}where~$\mathcal{N}\!\left( \mathbf{p}; \widehat{\boldsymbol{\mu}}_i^k, \widehat{\boldsymbol{\Sigma}}_i^k \right)$ is a 2D Gaussian density with weight~$\pi_i^k$ evaluated at~$\mathbf{p}\!=\!(\phi, \theta)$, with mean~$\widehat{\boldsymbol{\mu}}_i^k = (\hat{\mu}_{\mathrm{el}, i}^{k}, \hat{\mu}_{\mathrm{az}, i}^{k})$, and diagonal covariance~\textcolor{black}{$\widehat{\boldsymbol{\Sigma}}^k_i = \mathrm{diag}((\hat{\sigma}_{\mathrm{el}, i}^{k})^2, (\hat{\sigma}_{\mathrm{az}, i}^{k})^2)$}. 
The model is trained by minimizing the negative log-likelihood~$-\log p(\mathbf{p}\mid\Phi_i)$. Permutation-invariant training is employed to resolve source-label ambiguity. For calibration and testing, we follow~\cite{khurjekar2024multi} where the first and
second sources are identified according to their azimuth order. As in CRC-SSL-N, calibration is performed separately for each source. For the~$k$-th source, the nonconformity score for a candidate position~$\mathbf{p}\!=\!(\phi,\theta)$ is
\begin{equation}
s^{k}_i(\phi, \theta)= \sqrt{ \big( (\phi-\hat{\mu}_{\mathrm{el}, i}^{k}) / {\hat{\sigma}_{\mathrm{el}, i}^{k}} \big)^2 + \big( (\theta-\hat{\mu}_{\mathrm{az}, i}^{k}) / {\hat{\sigma}_{\mathrm{az}, i}^{k}} \big)^2 }.\notag
\end{equation}
Accordingly, the prediction region is expressed as
\begin{equation}
\mathcal{C}^{k}(\Phi_i)=\left\{(\phi,\theta):s_i^k(\phi, \theta) \le \hat{q}_{1-\alpha}^{k}\right\}.
\end{equation}
Geometrically,~$\mathcal{C}^{k}$ forms an ellipse centered at~$(\hat{\mu}_{\mathrm{el}, i}^{k},\hat{\mu}_{\mathrm{az}, i}^{k})$, whose semi-axes are~$\hat{q}_{1-\alpha}^{k}\hat{\sigma}_{\mathrm{el}, i}^{k}$ and~$\hat{q}_{1-\alpha}^{k}\hat{\sigma}_{\mathrm{az}, i}^{k}$, respectively.
}\looseness=-1

\begin{table}[!t]
\centering
\caption{`PT-SSL-N' results with~$\alpha_{\mathrm{MC}}\!=\!\delta\!=\!0.1$.} 
\scriptsize
\setlength{\tabcolsep}{1.2pt}

\resizebox{\columnwidth}{!}{%
\setlength{\tabcolsep}{3pt}
\begin{tabular}{@{} c *{12}{c} @{}}
\toprule
\multirow{4}{*}{\makecell{$T_{60}$\\[-1pt]{[ms]}}}
    & \multicolumn{6}{c}{SRP-PHAT}
    & \multicolumn{6}{c}{SRP-DNN} \\
\cmidrule(lr){2-7} \cmidrule(lr){8-13}
    & \multicolumn{2}{c}{$K=1$}
    & \multicolumn{2}{c}{$K=2$}
    & \multicolumn{2}{c}{$K=3$}
    & \multicolumn{2}{c}{$K=1$}
    & \multicolumn{2}{c}{$K=2$}
    & \multicolumn{2}{c}{$K=3$} \\
\cmidrule(lr){2-3} \cmidrule(lr){4-5} \cmidrule(lr){6-7}
\cmidrule(lr){8-9} \cmidrule(lr){10-11} \cmidrule(lr){12-13}
    & $P^{\mathrm{MC}}$ & PA\%
    & $P^{\mathrm{MC}}$ & PA\%
    & $P^{\mathrm{MC}}$ & PA\%
    & $P^{\mathrm{MC}}$ & PA\%
    & $P^{\mathrm{MC}}$ & PA\%
    & $P^{\mathrm{MC}}$ & PA\% \\
\midrule
400 & 0.08 & 0.31 & 0.10 & 23.19 & 0.05 & 44.80 
    & 0.09 & 0.12 & 0.06 & 0.22 & 0.05 & 5.07 \\
700 & 0.10 & 0.32 & 0.10 & 35.20 & 0.04 & 54.20 
    & 0.10 & 0.16 & 0.10 & 0.32 & 0.08 & 5.69 \\
\bottomrule
\label{tab:Pareto_known_srouce_count}
\end{tabular}}
\label{tab:mc_PA_srp_dnn_phat}
\end{table}

\printbibliography[title=References] 
\end{document}